\documentclass[twocolumn,prb,superscriptaddress,amsmath,amssymb,aps]{revtex4}
\usepackage{graphicx}
\usepackage{epsfig}
\usepackage{bm}
\usepackage{miller}
\usepackage{color}

\begin{document}

\title{BOPfox program for tight-binding and analytic bond-order potential calculations}

\author{T.~Hammerschmidt}
\affiliation{Atomistic Modelling and Simulation, ICAMS, Ruhr-Universit\"at Bochum, D-44801 Bochum, Germany}
\affiliation{Department of Materials, University of Oxford, Parks Road, Oxford OX1 3PH, United Kingdom}

\author{B.~Seiser}
\affiliation{Atomistic Modelling and Simulation, ICAMS, Ruhr-Universit\"at Bochum, D-44801 Bochum, Germany}
\affiliation{Department of Materials, University of Oxford, Parks Road, Oxford OX1 3PH, United Kingdom}

\author{M.~E.~Ford}
\affiliation{Atomistic Modelling and Simulation, ICAMS, Ruhr-Universit\"at Bochum, D-44801 Bochum, Germany}
\affiliation{Department of Materials, University of Oxford, Parks Road, Oxford OX1 3PH, United Kingdom}

\author{A.N.~Ladines}
\affiliation{Atomistic Modelling and Simulation, ICAMS, Ruhr-Universit\"at Bochum, D-44801 Bochum, Germany}

\author{S.~Schreiber}
\affiliation{Atomistic Modelling and Simulation, ICAMS, Ruhr-Universit\"at Bochum, D-44801 Bochum, Germany}

\author{N.~Wang}
\affiliation{Atomistic Modelling and Simulation, ICAMS, Ruhr-Universit\"at Bochum, D-44801 Bochum, Germany}

\author{J.~Jenke}
\affiliation{Atomistic Modelling and Simulation, ICAMS, Ruhr-Universit\"at Bochum, D-44801 Bochum, Germany}

\author{Y.~Lysogorskiy}
\affiliation{Atomistic Modelling and Simulation, ICAMS, Ruhr-Universit\"at Bochum, D-44801 Bochum, Germany}

\author{C.~Teijeiro}
\affiliation{High-Performance Computing in Materials Science, ICAMS, Ruhr-Universit\"at Bochum, D-44801 Bochum, Germany}

\author{M.~Mrovec}
\affiliation{Atomistic Modelling and Simulation, ICAMS, Ruhr-Universit\"at Bochum, D-44801 Bochum, Germany}

\author{M.~Cak}
\affiliation{Atomistic Modelling and Simulation, ICAMS, Ruhr-Universit\"at Bochum, D-44801 Bochum, Germany}

\author{E.~R.~Margine}
\affiliation{Department of Materials, University of Oxford, Parks Road, Oxford OX1 3PH, United Kingdom}
\affiliation{Department of Physics, Applied Physics and Astronomy, Binghamton University, State University of New York, Vestal, New York 13850, USA}

\author{D.~G.~Pettifor}
\affiliation{Department of Materials, University of Oxford, Parks Road, Oxford OX1 3PH, United Kingdom}

\author{R.~Drautz}
\affiliation{Atomistic Modelling and Simulation, ICAMS, Ruhr-Universit\"at Bochum, D-44801 Bochum, Germany}
\affiliation{Department of Materials, University of Oxford, Parks Road, Oxford OX1 3PH, United Kingdom}

\begin{abstract}
Bond-order potentials (BOPs) provide a local and physically transparent description of the interatomic interaction.
Here we describe the efficient implementation of analytic BOPs in the BOPfox program and library.
We discuss the integration of the underlying non-magnetic, collinear-magnetic and noncollinear-magnetic tight-binding models
that are evaluated by the analytic BOPs. 
We summarize the flow of an analytic BOP calculation including the determination of self-returning 
paths for computing the moments, the self-consistency cycle, 
the estimation of the band-width from the recursion coefficients, and the termination of the BOP expansion.
We discuss the implementation of the calculations of forces, stresses and magnetic torques with analytic BOPs. 
We show the scaling of analytic BOP calculations with the number of atoms and moments, present options
for speeding up the calculations and outline different concepts of parallelisation.
In the appendix we compile the implemented equations of the analytic BOP methodology and comments on the implementation.
This description should be relevant for other implementations and further developments of analytic bond-order potentials.
\end{abstract}

\maketitle

\section{Introduction}

A key requirement for reliable atomistic simulations is a robust description of the interatomic interaction. 
Density-functional theory (DFT) calculations provide a reliable treatment of the bond chemistry in many systems 
but the accessible length- and time-scales are limited due to the computational effort. 
Larger systems and/or longer time scales become accessible by coarse-graining the
electronic structure in DFT to the tight-binding (TB) approximation 
and further on to the analytic bond-order potentials (BOPs)~\cite{Pettifor-95,Drautz-06,Finnis-07,Drautz-11,Drautz-15}. 
This leads to a transparent and intuitive framework for modelling the interatomic interaction, 
including covalent bond formation, charge transfer and magnetism. 

The analytic BOPs~\cite{Drautz-06,Drautz-11} are closely related to the numerical BOPs~\cite{Horsfield-96} 
as discussed in Refs.~\cite{Hammerschmidt-09-NIC,Cak-13}.
Both have been applied in simulations of different materials, see Ref.~\cite{Hammerschmidt-09-IJMR} for an overview.
Here we describe our implementation of analytic BOPs in the software package BOPfox~\cite{BOPfox-url}.
BOPfox has already been used in several 
publications~\cite{Cak-13,Hammerschmidt-08-2,Chen-10,Seiser-11-2,Hammerschmidt-11,Hammerschmidt-12-1,Schablitzki-13,Cak-14,Drain-14,Ford-14,Teijeiro-15,Ford-15,Hammerschmidt-16,Teijeiro-16-1,Teijeiro-16-2,Teijeiro-17,Jenke-18}
and is being continuously extended and optimised.
We point out similarities of TB/BOP calculations and computations carried out using other electronic structure methods, and 
discuss the peculiarities of analytic BOPs in detail. 
This comprehensive description of the algorithmic framework of analytic BOPs should be of use for other implementations and further developments of analytic BOPs.

In Sec.~\ref{sec:flow} we outline the program flow of TB/BOP calculations in BOPfox.
Section~\ref{sec:performance} is devoted to the discussion of the performance with regard to scaling, speed-up and parallelisation.
The full set of equations that is evaluated during an analytic BOP calculation is compiled in the appendix with
details of the implementation and references to the original derivations.

\section{Program flow}
\label{sec:flow}

\subsection{Overview}

\begin{figure}[h!]
\begin{center}
\includegraphics[width=0.99\columnwidth]{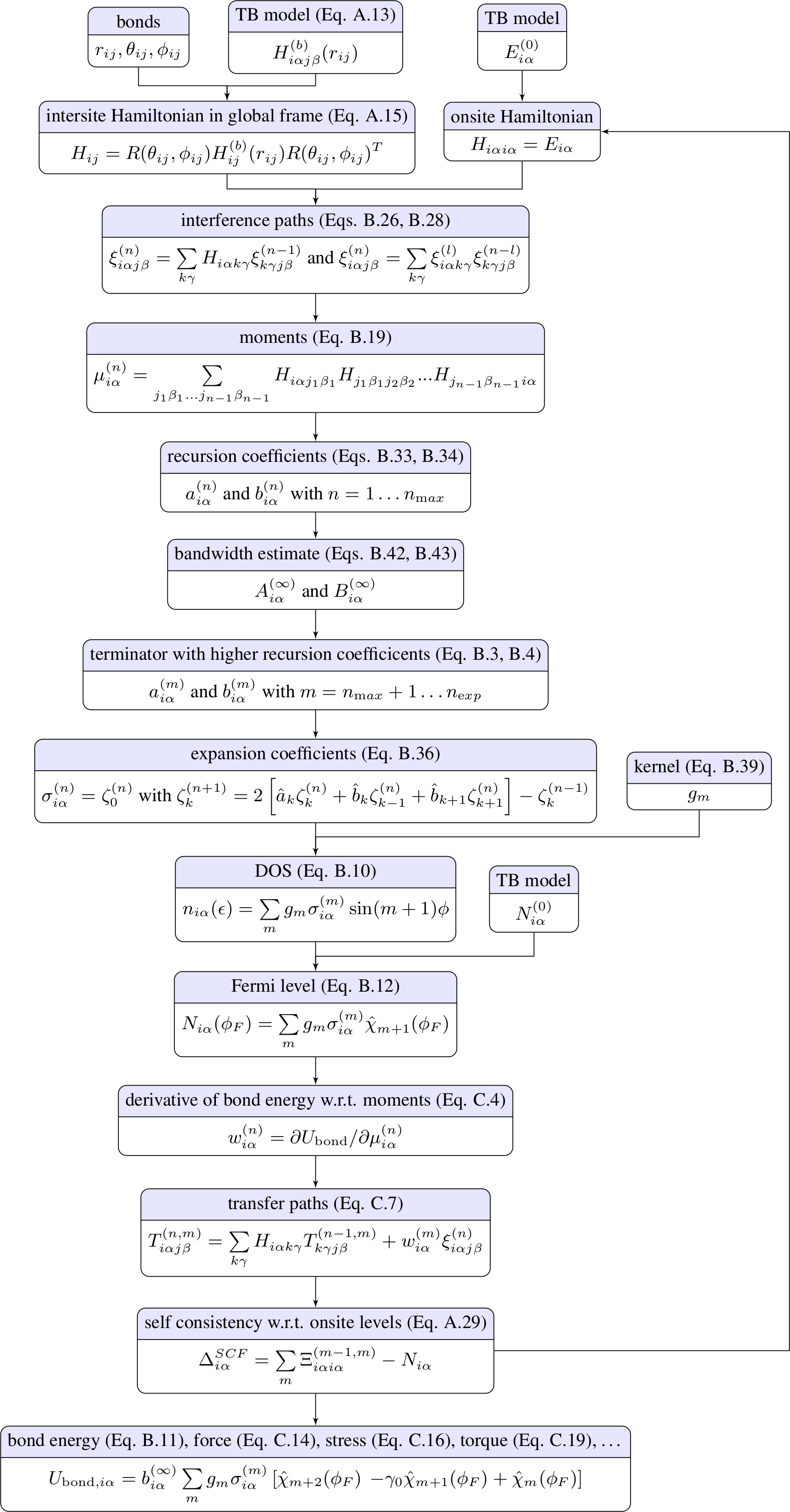}
\caption{\small Overview of the calculation of the bond energy for a non-magnetic system with analytic BOPs in BOPfox.}
\label{fig:BOP-calculation}
\end{center}
\end{figure}

The typical flow for computing the bond energy with a non-magnetic analytic BOP is sketched in Fig.~\ref{fig:BOP-calculation} and discussed in detail in the following.
The real-space BOP calculations can easily be complemented by reciprocal-space TB calculations that employ the same Hamiltonian matrix elements.

\subsection{Input files}

The initial stage of TB and BOP calculations in BOPfox is 
(i) reading the central control file (\verb+infox.bx+),
(ii) the specified structure file (default: \verb+structure.bx+) and 
(iii) the specified model file with the TB/BOP parameters (default: \verb+models.bx+).
The presently available TB/BOP models in BOPfox include parameters for magnetic calculations for
Fe~\cite{Ford-14,Mrovec-11,Madsen-11}, Fe-C~\cite{Hatcher-12,Schreiber-inprep},
for non-magnetic calculations for V~\cite{Lin-14}, Cr~\cite{Lin-14}, 
Nb~\cite{Cak-14,Lin-14}, Mo~\cite{Cak-14,Lin-14}, Ta~\cite{Cak-14,Lin-14}, W~\cite{Cak-14,Lin-14,Mrovec-07-2},
Ir~\cite{Cawkwell-06}, Si-N~\cite{Gehrmann-15}, and a canonical $d$-band model~\cite{Andersen-78}.
The set of TB/BOP parametrisations available in BOPfox is being constantly extended.

\subsection{Initialisation}

Two neighbour-lists of the crystal structure are created by setting up ghost cells and constructing cell linked-lists.
The implementation scales linearly with the number of atoms. The short-ranged neighbour-list is used for the construction of the intersite matrix elements
of the Hamiltonian ($H_{i\alpha j\beta}^{ }$ in Fig.~\ref{fig:BOP-calculation}), interference paths ($\xi_{i\alpha j\beta}^{(n)}$ in Fig.~\ref{fig:BOP-calculation})
and transfer paths ($T_{i\alpha j\beta}^{(n,m)}$ in Fig.~\ref{fig:BOP-calculation}). 
The second, long-range neighbour-list is used for the evaluation of the repulsive energy.

\subsection{Hamiltonian}

For each pair of atoms, the Hamiltonian matrix elements $H_{i\alpha j\beta}$ are constructed (Eq.~\ref{eq:Hij}) with the specified 
tight-binding model and rotated to the global coordinate system (Eq.~\ref{eq:Hij_rot}). 
TB/BOP calculations taking into account collinear or non-collinear magnetism use Hamiltonians with spin-dependent onsite levels 
as given in Eq.~\ref{eq:onsite_CM} and~\ref{eq:onsite_NCM}, respectively.
The implementation of collinear magnetism in BOPfox uses a loop over the $\uparrow$ and $\downarrow$ spin channels.
The calculations for the individual spin channels are very similar to non-magnetic BOP calculations.
The similar processes involved in non-collinear magnetic calculations, collinear magnetic calculations and non-magnetic calculations (see~\ref{sec:BOP_forces}) 
allow reuse of large portions of the code for each type of calculation. Switching the implementation to non-collinear magnetism is controlled by a 
preprocessor flag in the \verb+Makefile+ that includes the relevant parts of the source code.

\subsection{DOS and Fermi energy}

A key difference between the TB and BOP implementations is the calculation of the local density of states (DOS) $n_{i\alpha}(E)$:
(i) In analytic BOP calculations the pairwise $H_{i\alpha j\beta}$ are used to construct $n_{i\alpha}(E)$ in real space as outlined in~\ref{sec:DOS}.
(ii) In TB calculations the $H_{i\alpha j\beta}$ are used to generate a Hamiltonian with periodic boundary conditions that is diagonalised in reciprocal
space using LAPACK routines~\cite{Anderson-99}. 

The local DOS $n_{i\alpha}(E)$, whether obtained using TB calculations in reciprocal space or using BOP calculations in real space, is integrated up to the Fermi energy $E_{\rm{F}}$.
The Fermi energy is determined by the bisection method to match the sum of electrons in all orbitals with the total number of electrons in the system.

\subsection{Self-consistency}

The onsite levels $H_{i\alpha i\alpha}$ are optimised in the self-consistency loop (Eq.~\ref{eq:TB_SCF} or Eq.~\ref{eq:BOP_SCF}) until 
the contributions to the binding energy (Eqs.~\ref{eq:U_B}-~\ref{eq:TB-Ues}) and the forces (Eq.\ref{eq:F_k}) can be computed. 
The self-consistency condition in TB and BOP calculations is approached iteratively.
The onsite levels $E_{i\alpha}^{(n+1)}$ of step $n+1$ in the self-consistency loop are computed according to Eqs.~\ref{eq:TB_SCF} and~\ref{eq:BOP_SCF} 
from $n_{i\alpha}^{(n)}(E)$ that was obtained for the Hamiltonian with onsite levels $E_{i\alpha}^{(n)}$. 
With the new $E_{i\alpha}^{(n+1)}$, the Hamiltonian is updated and the new $n_{i\alpha}^{(n+1)}(E)$ is computed. 
In BOPfox, the input and output values of the onsite levels can be mixed (i) linearly, (ii) with the Broyden method~\cite{Broyden-65},
(iii) with the FIRE algorithm~\cite{Bitzek-06} or (iv) with molecular dynamics of onsite levels using a damped Verlet algorithm.
In all mixers, the self-consistency loop is carried out until the specified convergence limit or maximum number of steps is reached. 
The convergence of the different mixers depends on the particular system at hand, particularly for magnetic systems~\cite{Soin-11}.

\subsection{Energy and force contributions}

In TB, the bond energy is obtained by integrating the local electronic DOS $n_{i\alpha}(E)$ of the eigenvalues, which result from diagonalisation of the Hamiltonian, with  
the Methfessel-Paxton scheme~\cite{Methfessel-89} or the improved tetrahedron method~\cite{Bloechl-94}. 
In analytic BOPs, the bond energy is determined analytically from the local electronic DOS $n_{i\alpha}(E)$ and the Fermi energy $E_{\rm{F}}$, see~\ref{sec:BindingEnergy}.

In both TB and BOP calculations, the forces can be used for structural relaxation and MD simulations within BOPfox.
The current implementation includes several relaxation algorithms (e.g. damped MD, conjugate gradient~\cite{Gilbert-92},
L-BFGS~\cite{Zhu-94,Byrd-95}, FIRE~\cite{Bitzek-06}) as well as standard MD schemes (e.g. Verlet~\cite{Verlet-67}, velocity Verlet~\cite{Swope-82}).

\subsection{BOPfox as library: BOPlib}

BOPfox provides an application programming interface (API) for communication with external software.
The API takes the system configuration (species, positions, onsite levels, etc.) as arguments, starts a TB/BOP calculation
and returns atomic binding energies, forces, stresses and torques.
The combination of API and BOPfox subroutines can be compiled to a static or dynamic library called BOPlib. 
\begin{figure}[htb]
\begin{center}
\includegraphics[width=0.99\columnwidth]{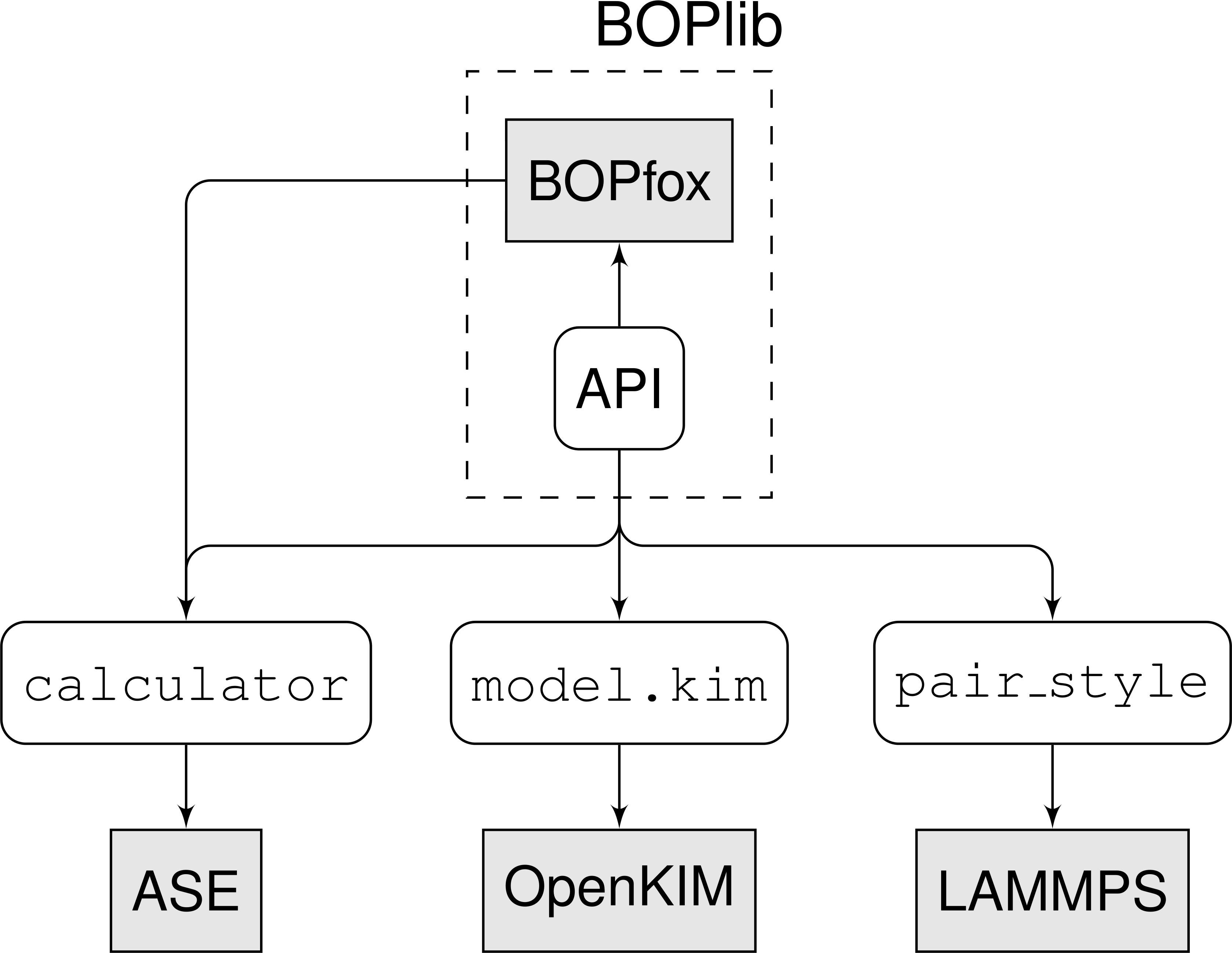}
\caption{\small Combination of BOPfox with ASE~\cite{Bahn-02}, openKIM~\cite{Tadmor-11} and LAMMPS~\cite{Plimpton-95} by the BOPlib API.}
\label{fig:BOPlib} 
\end{center}
\end{figure}
With BOPlib the TB/BOP calculations can be fully integrated with other external software as sketched in Fig.~\ref{fig:BOPlib}.
In particular, BOPfox can be addressed from ASE~\cite{Bahn-02} as {\tt calculator} with either BOPfox as system call or BOPlib as linked library.
BOPlib can also be configured as {\tt KIM model} to be linked to openKIM~\cite{Tadmor-11} and as {\tt pair\_style} potential to be linked with LAMMPS~\cite{Plimpton-95}.

\section{Performance}
\label{sec:performance}

\subsection{Scalability}

\begin{figure}[htb]
\begin{center}
\includegraphics[width=0.99\columnwidth]{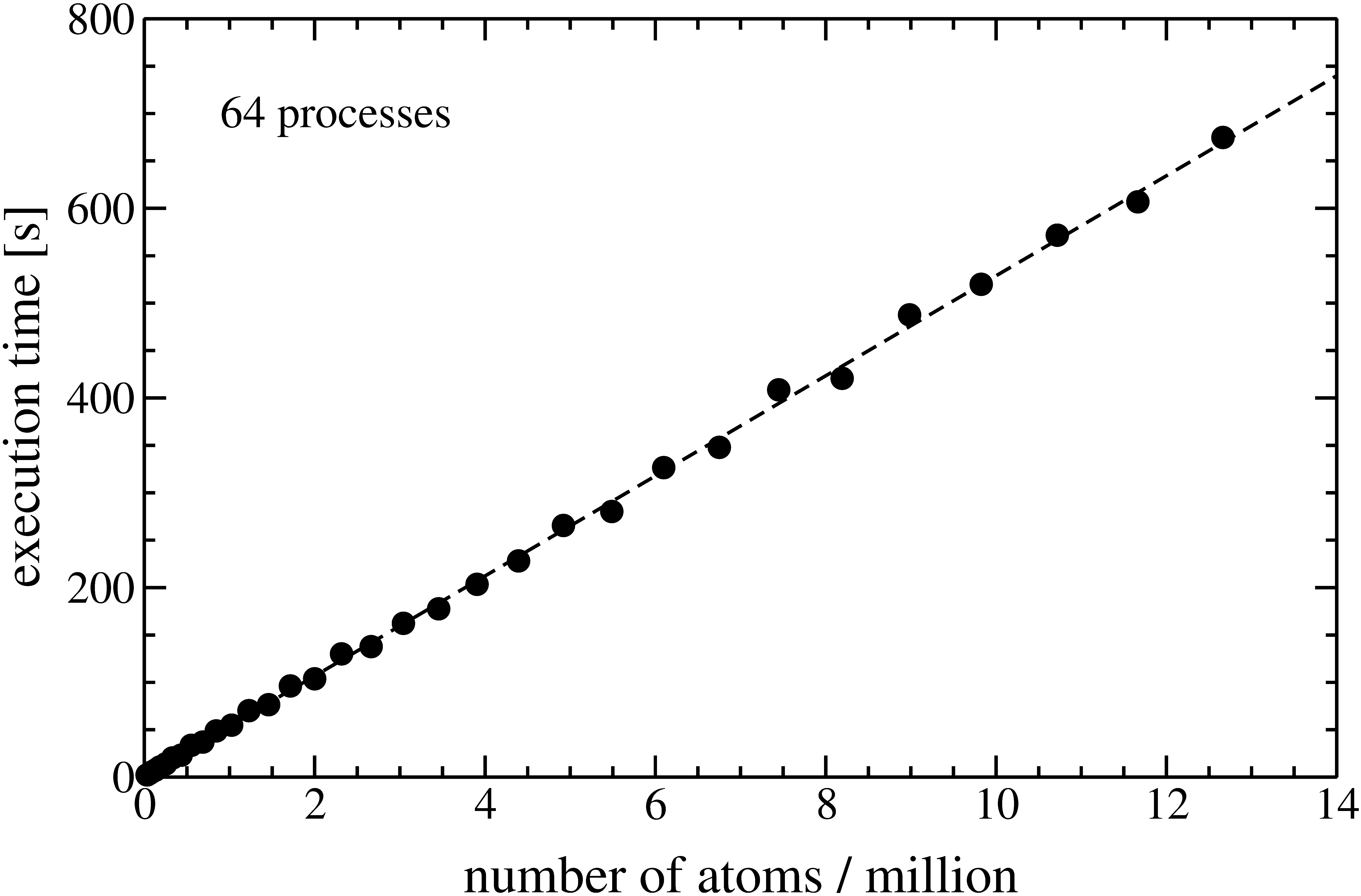}\
\caption{\small Linear scaling of the execution time with the number of atoms in the analytic BOP simulations.
The dashed line indicates a linear fit of the data points.
Technical details of the benchmark are given in Ref.~\cite{Teijeiro-16-2}.}
\label{fig:orderN} 
\end{center}
\end{figure}

The computational effort of energy and force calculations with analytic BOPs
is largely dominated by the evaluation of interference paths (Eq.~\ref{eq:momDOS})
and transfer paths (Eq.~\ref{eq:Xi}).
The theoretical scalability of the computational effort with respect to the
number of atoms and the number of moments is discussed in a detailed complexity 
analysis and systematic benchmarks in Ref.~\cite{Teijeiro-16-1}.
For typical choices of the number of moments, the complexity of the calculations 
increases with the number of moments to the power of approximately 4.5. The implementation 
of analytic BOPs in BOPfox reaches this theoretical scaling limit~\cite{Teijeiro-16-1}.
The increase in the computational effort with the number of atoms is linear (Fig.~\ref{fig:orderN}) 
due to the use of linear-scaling linked-cell lists and the locality of the BOP expansion.

\subsection{Speed-ups}

BOPfox provides several options to accelerate the energy and force calculations with analytic BOPs:

(i) The interference paths that are determined to evaluate the moments of the DOS are also
needed to compute the bond-order type term ${\tilde \Theta}_{i\alpha\nu j\beta\mu}$ for the
self-consistency (Eq.~\ref{eq:BOP_SCF}) and the forces (Eq.~\ref{eq:F_k}). An obvious approach
to improve the computational speed is therefore to store the interference paths. The resulting 
increase in memory limits this optimisation to moderate system sizes.

(ii) The self-consistency cycle involves the modification of onsite levels $E_{i\alpha}$ which 
necessitates the repeated computation of new interference paths (Eq.~\ref{eq:xi}).
This can hardly be avoided. However, small changes in the local atomic structure typically lead 
to only small changes in the self-consistent onsite levels. Hence for relaxations and MD 
simulations, the computation time can be reduced by initializing the onsite levels to the values 
of the previous step. For typical step sizes of relaxations or MD simulations,
this leads to significant speed-ups in successive self-consistent 
energy or force evaluations as fewer self-consistency steps need to be carried out.

(iii) In many cases the interatomic interaction is dominated by
the influence of the local environment of a given atom rather than effects due to atoms located further away. 
In the BOP framework, this expected short-sightedness of the interaction corresponds to a greater importance of the interference 
paths which sample the nearby environment as compared to those that reach out to more distant 
atoms. A straight-forward improvement in performance is, therefore, to introduce a maximum radius for the 
interference paths. In this way the immediate neighbourhood is fully sampled, while the paths that 
reach beyond a specified maximum radius are neglected. This introduces an additional level of
approximation.

\subsection{Parallelisation}

The computation of forces and energies using analytic BOPs is perfectly suited for parallel execution. 
BOPfox provides different concepts of parallelisation. Here we provide only an overview, the 
details and performance analysis are discussed in detail in the respective references given below.
Switching between different parallelisations is performed during compilation time with preprocessor flags.

(i) The shared-memory parallelisation based on OpenMP provides a straight-forward parallelisation
of the loops for computing the interference paths (Eqs.~\ref{eq:xi_rec}-\ref{eq:xi_merge}) and the 
transfer matrices (Eqs.~\ref{eq:Xi}-\ref{eq:Xi-merge}).
In this implementation all operations make use of the same arrays which are allocated for the whole 
simulation cell. Therefore the maximum size of the simulation cell is limited by the available
memory. 

(ii) The shared-memory parallelisation~\cite{Teijeiro-16-2} based on MPI uses a TODO list of operations
that is distributed to different threads. As for the shared-memory OpenMP parallelisation, the
working arrays are allocated for the whole simulation cell, which leads to a memory limitation.
This parallelisation approach is also suitable and implemented for GPU processing.

(iii) The distributed-memory parallelisation~\cite{Teijeiro-16-2} based on MPI performs a domain decomposition
of the simulation cell and thereby reduces the memory required per thread of the parallel execution. 
This implementation was optimised to reduce communication and to avoid redundant operations due to
the overlap of interference-paths calculations in the distributed domains.
\begin{figure}[htb]
\begin{center}
\includegraphics[width=0.99\columnwidth]{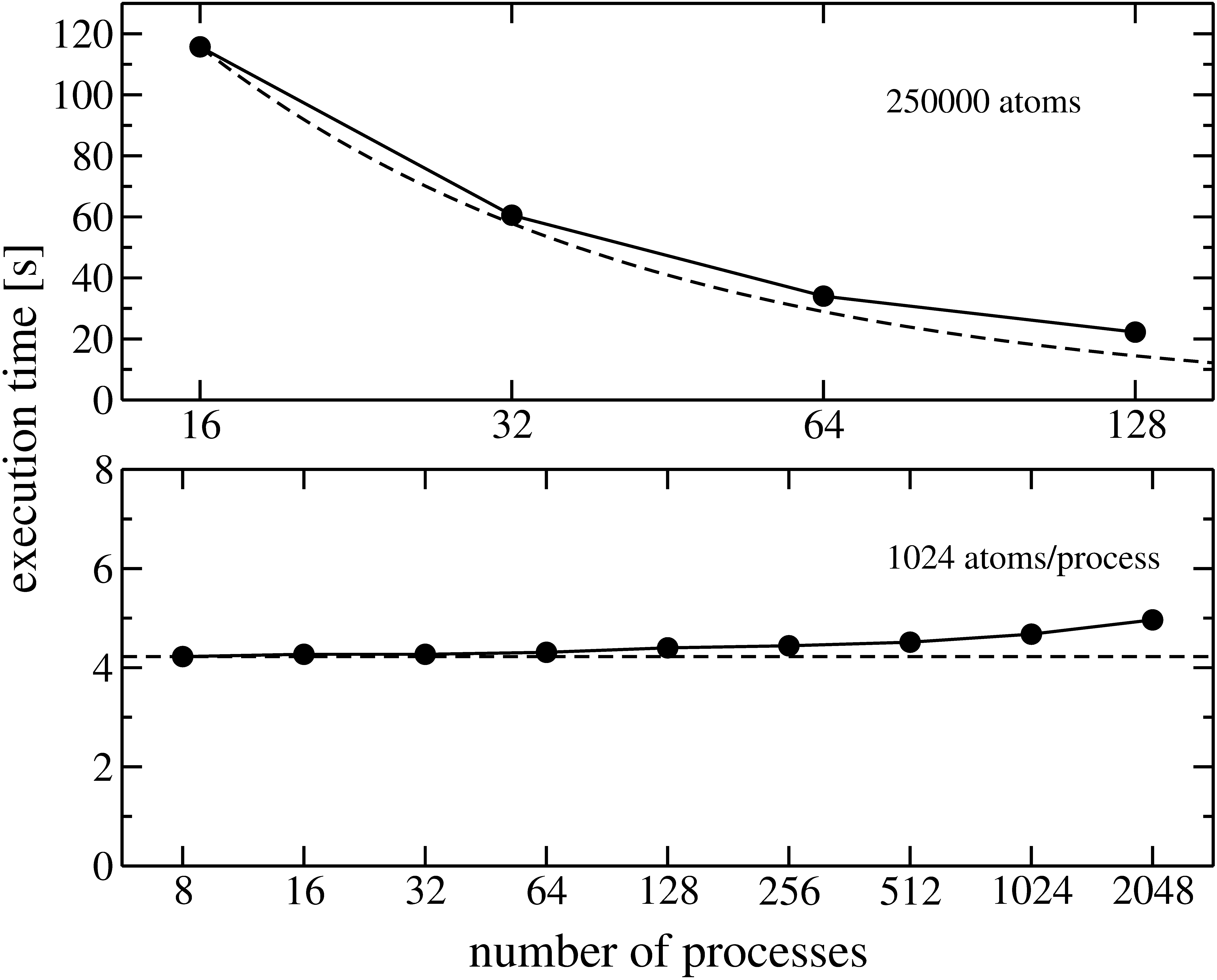}
\caption{\small Strong scaling of execution time with the number of processes for fixed system size (top) 
and weak scaling of execution time with the number of processes for fixed size of individual processes (bottom).
The dashed lines indicate ideal strong scaling and ideal weak scaling.
Technical details of the benchmark are given in Ref.~\cite{Teijeiro-16-2}.}
\label{fig:scaling} 
\end{center}
\end{figure}
The implementation in BOPfox reaches excellent strong scaling (Fig.~\ref{fig:scaling}, top), i.e. a linear decrease of the computation time 
for a fixed system size with the number of processes. At the same time it also shows excellent weak scaling (Fig.~\ref{fig:scaling}, bottom),
i.e. a constant execution time for increasing system size at a constant number of atoms per process.

(iv) The hybrid parallelisation~\cite{Teijeiro-17} is a combination of shared-memory and 
distributed-memory parallelisation that was developed to make use of the multi-core CPU architectures and 
multi-threading-capabilities of modern supercomputers. Here, the system is decomposed into domains that 
are distributed to different nodes using MPI. On each node the operations are then carried out on the 
same memory using OpenMP.

\section{Conclusions}

Analytic BOPs provide a local and physically transparent description of the interatomic interaction.
The BOPfox program package provides an implementation of analytic BOPs for non-magnetic, 
collinear-magnetic and noncollinear-magnetic calculations. It computes analytic forces, 
stresses and magnetic torques. 
For completeness, we compiled the implemented equations of the analytic BOPs with references to the original 
publications and comments on the implementation in the appendix.
This comprehensive description of the algorithmic framework should prove beneficial for a broader community of users and developers of analytic BOPs.

The implementation is highly efficient and provides linear scaling of the computation time for energies and 
forces with the number of atoms.
The different parallelisations make it possible to run the calculations with optimum use of the hardware resources for a 
given problem size.
The program can be compiled as standalone program or as library with an API for linking with an external software.

\section*{Acknowledgements}

We wish to dedicate this paper to the memory of our coauthor Professor David G. Pettifor CBE FRS, who sadly passed away before the work was completed.
We are grateful to Ting Qin, Paul Kamenski, Johnny Drain, Jan Gehrman, Aleksey Kolmogorov, Thomas Schablitzki, Martin Staadt and Jutta Rogal for discussions and for their feedback using BOPfox.
TH, BS, RD, and DGP acknowledge funding from the Engineering and Physical Sciences Research Council (EPSRC) of the United Kingdom through the project {\em Alloys by Design}.
AL, TH and RD acknowledge financial support by the German Research Foundation (DFG) through research grant HA 6047/4-1 and project C1 and C2 of the collaborative research centre SFB/TR 103.
SS, MC, TH and RD acknowledge funding through the project {\em Damage Tolerant Microstructures in Steel} by thyssenkrupp Steel Europe AG and Benteler Steel Tube GmbH.
MEF acknowledges funding from the EPSRC through a University of Oxford, Department of Materials Doctoral Training Award (DTA).  
CT acknowledges funding from thyssenkrupp Steel Europe AG through the HPC group.
MC acknowledges financial support by the DFG through research grant CA 1553/1-1.
E.R.M. acknowledges the NSF support (Award No. OAC-1740263).
Part of the work of NW and AL  was carried out in the framework of the International Max-Planck Research School SurMat.

\bibliographystyle{elsarticle-num}
\bibliography{bibliography}

\appendix

\section{Binding energy in TB and BOP}
\label{sec:BindingEnergy}
\setcounter{figure}{0}

\subsection{Energy contributions}

The TB and BOP calculations within BOPfox are based on the TB bond model~\cite{Drautz-15,Sutton-88} that can be obtained as a second-order expansion
of the DFT energy~\cite{Drautz-11}. In the absence of external fields the total binding energy is given by
\begin{equation}
 \label{eq:U_B}
  U_B  = U_{\rm{bond}} + U_{\rm{prom}} + U_{\rm{ion}} + U_{\rm{es}} + U_{\rm{rep}} + U_{\rm{X}}\, .
\end{equation}
The covalent bond energy $U_{\rm{bond}}$ summarizes the energy that originates from the formation of chemical bonds between the atoms.
Its onsite representation 
\begin{equation}
 \label{eq:TB-Ubond}
 U_{\rm{bond}} = \sum\limits_{i\alpha\nu} \int\limits^{E_{\rm{F}}} \left(E -E_{i\alpha\nu} \right) n_{i\alpha\nu}(E) {\rm d}E 
\end{equation}
is the integral of the local electronic DOS $n_{i\alpha\nu}(E)$ up to the Fermi energy $E_{\rm{F}}$ for each orbital $\alpha$ and spin $\nu$ of atom $i$
with onsite level $E_{i\alpha\nu}$.
The equivalent intersite representation
\begin{equation}
 \label{eq:TB-Ubond-intersite}
 U_{\rm{bond}} = \sum\limits_{i\alpha\nu j\beta\mu}^{i\alpha\nu \ne j\beta\mu} \beta_{i\alpha\nu j\beta\mu} n_{j\beta\mu i\alpha\nu}
\end{equation}
is expressed in terms of the density-matrix elements $n_{i\alpha\nu j\beta\mu}$ (Eq.~\ref{eq:nialphajbeta})
that are identical to the bond order $\Theta_{i\alpha\nu j\beta\mu}(\phi_F)$ aside from a factor of two for non-magnetic systems.
The bond integrals~\cite{Drautz-11,Margine-14}
\begin{equation}
 \label{eq:beta_ij} 
 \beta_{i\alpha\nu j\beta\mu} = H_{i\alpha\nu j\beta\mu} - \frac{1}{2}\left(E_{i\alpha\nu}+E_{j\beta\mu} \right) S_{i\alpha\nu j\beta\mu}
\end{equation}
include the Hamiltonian matrix elements $H_{i\alpha\nu j\beta\mu}$ and overlap matrix elements $S_{i\alpha\nu j\beta\mu}$~\cite{Drautz-15}.
The promotion energy $U_{\rm{prom}}$ accounts for the redistribution of electrons across orbitals upon bond formation.
It is given by
\begin{equation}
 \label{eq:TB-Uprom} 
 U_{\rm{prom}} = \sum\limits_{i\alpha\nu} E_{i\alpha\nu}^{(0)} \left( N_{i\alpha\nu}^{ } - N_{i\alpha\nu}^{(0)}\right)
\end{equation}
with $(0)$ indicating the non-magnetic free atom as reference and the number of electrons 
\begin{equation}
 \label{eq:N} 
 N_{i\alpha\nu} =  \int\limits^{E_{\rm{F}}} n_{i\alpha\nu}(E) {\rm d}E \, .
\end{equation}
The deviation from charge-neutral atoms upon bond formation leads to charges 
\begin{equation}
q_{i\alpha\nu} = N_{i\alpha\nu}^{ }-N_{i\alpha\nu}^{(0)} \, .
\end{equation}
The energies associated with charge redistribution are approximated to depend only on the total atomic charge
\begin{equation}
q_i = \sum\limits_{\alpha\nu} q_{i\alpha\nu} \, .
\end{equation}
The energy to charge an atom is given by the onsite ionic energy
\begin{equation}
 \label{eq:TB-Uion} 
 U_{\rm{ion}} = \bar{E}_i q_i + \frac{1}{2} \sum\limits_{i} J_{ii} q_i^2
\end{equation}
that is determined by the electronegativity $\bar{E}_i$ and the resistance against charge transfer $J_{ii}$ 
that is related to the Hubbard U~\cite{Hubbard-63}.
The energy $\bar{E}_i q_i$ is obtained by a weighted average of the reference onsite levels~\cite{Drautz-11}
\begin{equation}
\label{eq:ebar}
\bar{E}_i = \sum_{\alpha} E_{i \alpha}^{(0)} \Delta q_{i \alpha}
\end{equation}
where $\Delta q_{i \alpha}$ is the amount of charge which is gained or lost by orbital $i\alpha$ due to minimization of the binding energy $U_B$.
The interaction of the charged atoms is given by the intersite electrostatic energy 
\begin{equation}
 \label{eq:TB-Ues} 
 U_{\rm{es}} = \frac{1}{2} \sum\limits_{ij}^{i\ne j} J_{ij}q_iq_j
\end{equation}
with the Coulomb parameter $J_{ij}$.
The repulsive energy $U_{\rm{rep}}$ includes all further terms of the second-order expansion of DFT~\cite{Drautz-15}
and is usually parametrised by empirical functions.
The exchange energy $U_{\rm{X}}$ due to magnetism is approximated by the typically dominating onsite contributions
\begin{equation}
 \label{eq:TB-UX} 
 U_{\rm{X}} = -\frac{1}{4} \sum\limits_i I_{i} m_i^2
\end{equation}
with $m_i$ the magnetic moment and $I_{i}$ the Stoner exchange parameter of atom $i$.
The preparation energy (Eq.~92 in Ref.~\cite{Drautz-15}) vanishes in an unscreened calculation.
Further contributions to the energy due to external magnetic or electric fields can be included~\cite{Drautz-11}.

\subsection{Hamiltonian}

\subsubsection{Construction}

For each interacting pair of atoms $i$ and $j$ with orbitals $\alpha$ and $\beta$, 
the structure of the pairwise Hamiltonian $H_{i j}^{(b)}$ in the coordinate system of the bond is given by
\begin{equation}
\label{eq:Hij}
\hspace{-0.3cm}
H_{i j}^{(b)}  = \begin{array}{ll} 
 & 
 \begin{array}{lllllllll} 
   js      &        & jp          &           &         &           & jd         &           &          \\
 \end{array}\\
 \begin{array}{l} 
  is\\
   \\
  ip\\
   \\
   \\
   \\
  id\\
   \\
   \\
 \end{array} 
& \hspace{-0.3cm}
 \left( \begin{array}{l|lll|lllll} 
  \sigma & \sigma & 0         & 0         & \sigma  & 0         & 0         & 0         & 0        \\
  \hline
  \sigma & \sigma & 0         & 0         & \sigma  & 0         & 0         & 0         & 0         \\
  0      & 0      & \pi       & 0         & 0       & \pi       & 0         & 0         & 0         \\
  0      & 0      & 0         & \pi       & 0       & 0         & \pi       & 0         & 0         \\
  \hline
  \sigma & \sigma & 0         & 0         & \sigma  & 0         & 0         & 0         & 0         \\
  0      & 0      & \pi       & 0         & 0       & \pi       & 0         & 0         & 0         \\
  0      & 0      & 0         & \pi       & 0       & 0         & \pi       & 0         & 0         \\
  0      & 0      & 0         & 0         & 0       & 0         & 0         & \delta    & 0         \\
  0      & 0      & 0         & 0         & 0       & 0         & 0         & 0         & \delta    \\
\end{array} \right)
\end{array}
\end{equation}
for the general case of an $spd$-valent atom $i$ interacting with an $spd$-valent atom $j$.
The superscript $(b)$ indicates the coordinate system of the bond aligned along the $z$ axis with ordering of the $p$ and $d$ orbitals as $p_z,p_x,p_y$ and $d_{3z^2-r^2}, d_{zx}, d_{yz}, d_{x^2-y^2}, d_{xy}$, respectively.
The values of the matrix elements $\sigma$ and $\pi$ differ in general for different combinations of orbitals (e.g. $\sigma(is,js) \ne \sigma(ip,jp)$) and atoms (e.g. $\sigma(ip,jd) \ne \sigma(id,jp)$).
For combinations of atoms with fewer types of valence orbitals, the Hamiltonian reduces accordingly.
The values of the matrix elements $H_{i\alpha j\beta}^{(b)}$ are determined for the interatomic distance $r_{ij} = |{\mathbf r}_{ij}| = |\mathbf{r}_{i}-\mathbf{r}_{j}|$ from the values 
of the distance-dependent bond integrals $\beta_{i\alpha j\beta}(r_{ij})$.
The functional form of $\beta_{i\alpha j\beta}(r_{ij})$ depends on the specific TB/BOP model and is, for example, power-law, exponential, or Goodwin-Skinner-Pettifor~\cite{Goodwin-89} type.
The interaction range can be smoothly forced to zero at $r_{\rm{cut}}$ by multiplication of $\beta_{i\alpha j\beta}(r_{ij})$ with a cosine function
\begin{equation}
\label{eq:cutoff}
 f_{\rm{cut}}(r_{ij}) = \frac{1}{2}\left({\rm cos}\left(\pi\left[\frac{r_{ij}-(r_{\rm{cut}} - d_{\rm{cut}})}{d_{\rm{cut}}}\right]\right)+1\right)
\end{equation}
for $r_{\rm{cut}}$ - $d_{\rm{cut}} \le r_{ij} \le r_{\rm{cut}}$. 
For each bond, the pairwise Hamiltonian initialised in the bond coordinate system is rotated to the global coordinate system 
\begin{equation}
\label{eq:Hij_rot}
H_{i j} = R(\theta_{ij},\phi_{ij}) H_{i j}^{(b)}(r_{ij}) R(\theta_{ij},\phi_{ij})^T
\end{equation}
using rotation matrices $R(\theta_{ij},\phi_{ij})$ with polar and azimuthal angles $\theta_{ij}$ and $\phi_{ij}$ determined from the orientation of the bond ${\mathbf r}_{ij}$ in the global coordinate system (see~\ref{sec:rot_mat}).

\subsubsection{Magnetism}
\label{sec:magnetism}

Magnetism enters the Hamiltonian $H_{i\alpha\mu j\beta\nu}$ via the explicit spin-dependence of the onsite levels 
$E_{i\alpha\mu\nu}$~\cite{Drautz-11}.
The spin indices $\mu$ and $\nu$ span the four quadrants of neighbouring electron spin $\uparrow\uparrow$, $\uparrow\downarrow$, $\downarrow\uparrow$ and $\downarrow\downarrow$.
The global onsite-level matrix of orbitals $\alpha$ of atom $i$~\cite{Ford-15}
\begin{equation}
 {\bm E}_{i\alpha} = \left( 
  \begin{array}{ll}
    E_{i\alpha\uparrow\uparrow}    \, \, E_{i\alpha\uparrow\downarrow}  \\
    E_{i\alpha\downarrow\uparrow}  \, \, E_{i\alpha\downarrow\downarrow}   
  \end{array} 
  \right)
\end{equation}
with onsite levels~\cite{Ford-14}
\begin{eqnarray}\label{eq:onsite_CM}
 E_{i\alpha\mu\nu} &=& H_{i\alpha\mu i\alpha\nu} \\
                   &=& H_{i\alpha i\alpha}^{(0)} \delta_{\mu\nu} + \mathbf {B}_i \cdot {\bm \sigma}_{\mu\nu}
                       - \frac{1}{2} I_i \mathbf{m}_i \cdot {\bm \sigma}_{\mu\nu} + J_i q_i \nonumber
\end{eqnarray}
depends on the non-magnetic onsite levels $H_{i\alpha i\alpha}^{(0)}$, any external magnetic field $\mathbf{B}_i$, the Pauli matrices ${\bm \sigma}_{\mu\nu}$,
the Stoner exchange integral $I_i$~\cite{Stoner-39} and the charge $q_i$.

In the case of \emph{collinear magnetism}~\cite{Ford-14} with identical axis of spin quantization for all atoms the magnetic moments are parallel or 
antiparallel to one another. In this case the global magnetic moment direction can be taken to lie along the z-axis of the unit cell.
Then the $\uparrow\downarrow$ and $\downarrow\uparrow$ modifications to $H_{i\alpha j\beta}^{(b)}$ vanish
and the global onsite-level matrix takes a diagonal form with decoupled $\uparrow\uparrow$ and $\downarrow\downarrow$ modifications.
Therefore, we may use separate $\uparrow$ and $\downarrow$ spin channels $\nu$ with onsite elements
\begin{equation}\label{eq:onsite_NCM}
 E_{i\alpha\nu} = H_{i\alpha i\alpha}^{(0)} - (-1)^\nu B_z + \frac{1}{2} (-1)^\nu I_i m_i + J_i q_i
\end{equation}
for a magnetic moment of
\begin{equation}
  m_{i} = \sum\limits_{\alpha} \left( N_{i\alpha\uparrow} - N_{i\alpha\downarrow} \right) \, .
\end{equation}

In the case of \emph{non-collinear magnetism}~\cite{Ford-15}, the axis of spin quantization is different for different atoms $i$.
However, with a unitary transformation ${\bm U}_{i\alpha}$, the diagonal form of ${\bm E}_{i\alpha}$ can be enforced 
\begin{equation}
\label{eq:onsite-noncollinear}
  {\bm E}_{i\alpha}^{{\rm (local)}} = {\bm U}_{i\alpha}^{ } {\bm E}_{i\alpha}^{ } {\bm U}_{i\alpha}^\dagger = 
    \left( \begin{array}{cc}
    E_{i\alpha}^{\uparrow\rm{(local)}}    & \, \, 0  \\
    0                                     & \, \, E_{i\alpha}^{\downarrow\rm{(local)}}   
  \end{array} 
  \right)
\end{equation}
by a rotation into a local coordinate system that is oriented along the local magnetic moment.
The transformation matrix ${\bm U}_{i\alpha}^{ }$ is defined~\cite{Kuebler-88} in terms of the angle $\alpha$ between the $z$ direction in the global space, ${\bm s}_z$, and the direction of the local magnetic moment, ${\bm s}_{i\alpha}$
\begin{equation}
\cos(\alpha) = {\bm s}_z \cdot {\bm s}_{i\alpha} 
\end{equation} 
and a vector ${\bm n}_{i\alpha}$ that is orthogonal to ${\bm s}_z$ and ${\bm s}_{i\alpha}$.
A computationally convenient way to express the transformation matrix is~\cite{Ford-15}
\begin{equation}
 {\bm U}_{i\alpha}^{ } = \cos\left( \frac{\alpha}{2} \right) {\bm 1} - i \left( {\bm \sigma} \cdot {\bm n}_{i\alpha} \right) \sin \left( \frac{\alpha}{2} \right)
\end{equation}
with the identity matrix ${\bm 1}$ and the vector of Pauli spin matrices ${\bm \sigma}$.

\subsubsection{Screening}

The analytic BOP calculations in BOPfox employ orthogonal TB models that can be
obtained by approximate transformations of non-orthogonal TB models.
This transformation leads to an environment dependency of the bond integrals $\beta_{i\alpha j\beta}(r_{ij})$ 
in the orthogonal TB model~\cite{NguyenManh-00} in terms of screening by environment 	atoms $k$ with orbitals $\gamma$. 

The transformation to an orthogonal basis is achieved by a L{\"o}wdin transformation~\cite{Loewdin-50}
\begin{equation}
 \tilde{H}_{i\alpha j\beta} = S_{i\alpha k\gamma}^{-1/2}  H_{k\gamma l\delta}^{ } S_{l\delta j\beta}^{-1/2} \, .
\end{equation}
The diagonal elements of the overlap matrix are one, it can therefore be written as
\begin{equation}
   S_{i\alpha j\beta} = \delta_{i\alpha j\beta} + O_{i\alpha j\beta} \,.
\end{equation}
where $O_{ij} = S_{ij}$ for $i\ne j$ and zero otherwise.
Similarly, we can write
\begin{equation}
   S_{i\alpha j\beta}^{-1/2} = \delta_{i\alpha j\beta} - \frac{1}{2} {\mathfrak S}_{i\alpha j\beta}  \, .
\end{equation}
The screened orthogonal Hamiltonian matrix elements are given as~\cite{Drautz-15}
\begin{eqnarray}
  \tilde{H}_{i\alpha j\beta}^{(0)} 
    &=& H_{i\alpha j\beta}^{(0)}  \nonumber \\
    && - \frac{1}{2}\left( H_{i\alpha k\gamma}^{(0)} {\mathfrak S}_{k\gamma j\beta} + {\mathfrak S}_{i\alpha k\gamma} H_{k\gamma j\beta}^{(0)} \right) \nonumber \\
    && + \frac{1}{4} {\mathfrak S}_{i\alpha k\gamma} H_{k\gamma l\delta}^{(0)} {\mathfrak S}_{l\delta j\beta} 
\end{eqnarray}
where the bond between atoms $i$ and $j$ is screened by atom $k$.
The matrices $O_{i\alpha j\beta}$ are constructed analogously to the Hamiltonian (Eq.~\ref{eq:Hij}) 
with pairwise distance-dependent parametrisations.
In BOPfox, the screening is implemented up to the linear term in ${\mathfrak S}$,
while ${\mathfrak S}$ is approximated to first order as
\begin{equation}
   {\mathfrak S}_{i\alpha j\beta} = O_{i\alpha j\beta} \, .
\end{equation}

\subsection{Self-consistency}
\label{sec:SCF}

The onsite levels $E_{i\alpha}$ of the different atoms $i$ in the system are optimised in a self-consistency loop 
in order to minimise the binding energy (Eq.~\ref{eq:U_B}).
The target quantity $\Delta_{i\alpha}^{SCF}$ that is minimised with respect to onsite levels~\cite{Drautz-11}, defined by  
\begin{equation}
  \frac{\partial U_B}{\partial E_{i\alpha}}=\Delta_{i\alpha}^{SCF} \rightarrow 0
\end{equation}
can be expressed for the case of BOP calculations as 
\begin{equation}
  \label{eq:BOP_SCF}
  \Delta_{i\alpha}^{SCF} = \tilde{\Theta}_{i\alpha i\alpha} - N_{i\alpha}
               = \sum\limits_{m} \Xi_{i\alpha i\alpha}^{(m-1,m)} - N_{i\alpha} \, .
\end{equation}
The bond-order like term $\Xi_{i\alpha i\alpha}^{(m-1,m)}$ that includes gradients of the moments with respect to onsite levels 
is explained in detail in~\ref{sec:BOP_forces}.

The corresponding minimisation target for TB calculations can be written as 
\begin{eqnarray}
  \label{eq:TB_SCF}
  \Delta_{i\alpha}^{SCF} &=& E_{i\alpha} - \left( E_{i\alpha}^{(0)} + \sum\limits_{j\beta} J_{i\alpha j\beta} q_j \right) \\ \nonumber
               &=& \Delta E_{i\alpha} - \sum\limits_{j\beta} J_{i\alpha j\beta} q_j \, .
\end{eqnarray}

Local-charge neutrality can be enforced by the alternative target quantity
\begin{equation}
  \label{eq:BOP_LCN}
  \Delta_{i\alpha}^{SCF} = N_{i\alpha}^{(0)} - N_{i\alpha}
\end{equation}
or, implicitly, by large values of $J_{i\alpha i\alpha}$.

For \emph{non-collinear magnetism}~\cite{Ford-15}, the gradient of the binding energy with respect to local onsite levels 
${\bm E}_{i\alpha\nu}^{{\rm (local)}}$ (Eq.~\ref{eq:onsite-noncollinear}), i.e.,
\begin{equation}
   \frac{\partial U_B}{\partial E_{i\alpha\nu}^{\rm (local)}}= {\bm \tilde{\Theta}}_{i\alpha\nu i\alpha\nu}^{\rm (local)} - N_{i\alpha\nu}
\end{equation}
involves the unitary transformation
\begin{equation}
{\bm \tilde{\Theta}}_{i\alpha\nu i\alpha\nu}^{\rm (local)} = {\bm U}_{i\alpha\nu}^{ } {\bm \tilde{\Theta}}_{i\alpha\nu i\alpha\nu}^{ } {\bm U}_{i\alpha\nu}^\dagger  
\end{equation}

\section{Bond energy in analytic BOPs}
\label{sec:BondEnergyBOP}
\setcounter{figure}{0}    

\subsection{Density of states}
\label{sec:DOS}

\begin{figure}[htb]
\begin{center}
\includegraphics[width=0.99\columnwidth]{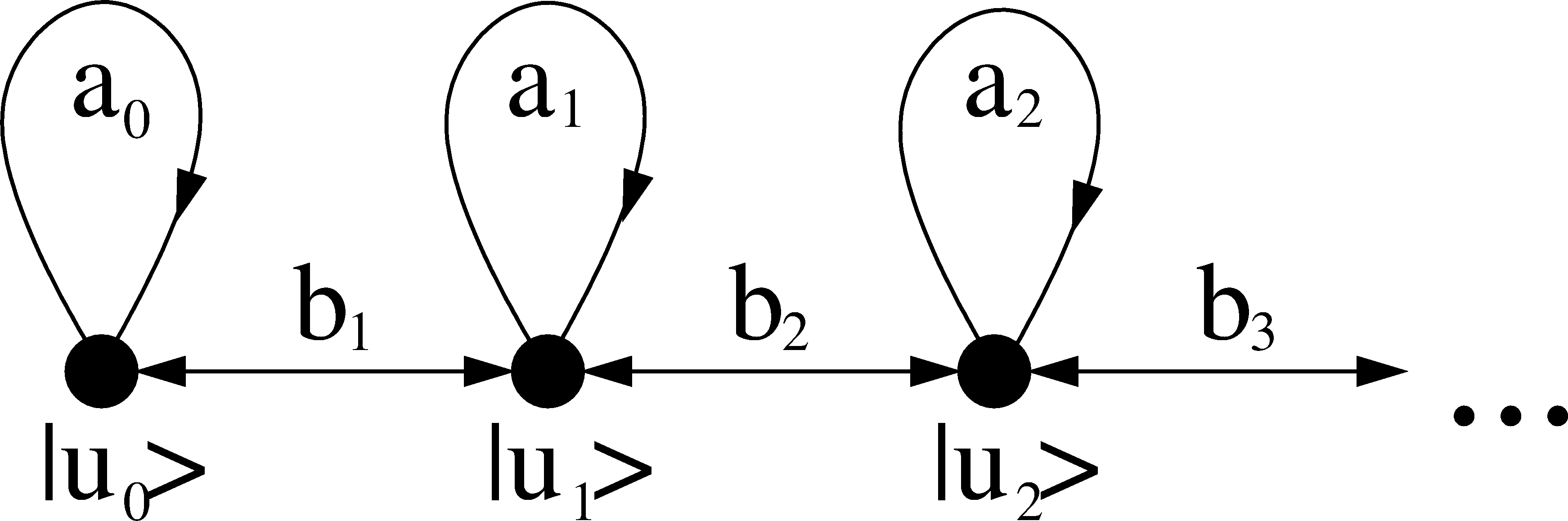}
\caption{\small Graphical representation of the recursion Hamiltonian as a one-dimensional chain: the Lanczos chain.}
\label{fig:linear-chain} 
\end{center}
\end{figure}

In analytic BOPs, the local density of states $n_{i\alpha}(\epsilon)$ required for the calculation of the bond energy (Eq.~\ref{eq:TB-Ubond}),
\begin{equation}\label{eq:nialpha}
n_{i\alpha}(\epsilon) = \frac{2} {\pi} \sqrt{1-\epsilon^{2}} \sum\limits_m g_m \sigma_{i\alpha}^{(m)} P_{m}(\epsilon)
\end{equation}
is determined analytically~\cite{Drautz-06,Drautz-11,Drautz-15,Seiser-13} 
using Chebyshev polynomials of the second kind $P_{m}(\epsilon)$ (see~\ref{sec:chebyshev}),
structure-dependent expansion coefficients $\sigma_{i\alpha}^{(m)}$ (see~\ref{sec:sigma}), and
damping factors $g_m$ (see~\ref{sec:g_m}).
The expansion of the DOS is based on a transformation of the Hamiltonian to a tridiagonal form~\cite{Haydock-80-1}
\begin{displaymath}
\label{eq:Hmatrix}
\langle u_n | \hat{H} | u_m \rangle = 
\left( \begin{array}{ccccccc}
a^{(0)} \, & b^{(1)}    &            &            &          &          \\
b^{(1)}    & a^{(1)} \, & b^{(2)}    &            &          &          \\
           & b^{(2)}    & a^{(2)} \, & b^{(3)}    &          &          \\
           &            & b^{(3)}    & a^{(3)} \, & \ddots    &          \\
           &            &            & \ddots     & \ddots \, & \ddots    \\
           &            &            &            & \ddots    & \ddots \, 
\end{array} \right) 
\end{displaymath}
with all other entries identical to zero. 
This Hamiltonian corresponds to a one-dimensional chain with only nearest-neighbour matrix elements, see Fig.~\ref{fig:linear-chain}.
that can be solved by recursion~\cite{Haydock-72} using the Lanczos algorithm~\cite{Lanczos-50}
to obtain the local DOS
\begin{eqnarray}
\label{eq:DOS-G}
n_{i\alpha}(E)  =  -\frac{1}{\pi} \mathrm{Im}\, \cfrac{1}{E-a_{i\alpha}^{(0)} -\cfrac{{b_{i\alpha}^{(1)}}^2}{E-a_{i\alpha}^{(1)}-\cfrac{{b_{i\alpha}^{(2)}}^2}{\ddots}}}
\end{eqnarray}
in terms of the recursion coefficients $a_{i\alpha}^{(m)}$ and $b_{i\alpha}^{(m)}$.
In practice, the recursion is terminated at some level $n$ by making assumptions for the values of $a_{i\alpha}^{(m)}$ and $b_{i\alpha}^{(m)}$ for $m>n$.
This corresponds to taking the energy calculation to a local scheme which requires convergence with respect to $n$.
In BOPfox the required recursion coefficients $a_{i\alpha}^{(m)}$ and $b_{i\alpha}^{(m)}$ for $m>n$ can be taken (i) as constant, (ii) as weighted average and (iii) as oscillating.

Taking the recursion coefficients as constant values
\begin{equation}
\label{eq:abinf}
a_{i\alpha}^{(m)}=a_{i\alpha}^{(\infty)}, \quad b_{i\alpha}^{(m)}=b_{i\alpha}^{(\infty)} \qquad \mathrm{for} \, \, m>n 
\end{equation}
corresponds to the so-called square-root terminator as the tail of the continued fraction 
can then be given analytically as a square-root function~\cite{Haydock-75-2}.
The different approaches to obtain the values of the asymptotic recursion coefficients $a_{i\alpha}^{(\infty)}$ and $b_{i\alpha}^{(\infty)}$ in BOPfox are summarized in~\ref{sec:terminator}. 

Taking $a_{i\alpha}^{(m)}$ and $b_{i\alpha}^{(m)}$ for $m>n$ as weighted averages~\cite{Ford-15} over $m_{\textrm{rec}}^{\textrm{max}}$ recursion levels 
\begin{equation}
\label{eq:weighted}
a_{i\alpha}^{(\textrm{approx})} = \frac{\sum\limits_{m=0}^{m_{\textrm{rec}}^{\textrm{max}}}w_m a_{i\alpha}^{(m)}}{\sum\limits_{m=0}^{m_{\textrm{rec}}^{\textrm{max}}} w_m} , \quad
b_{i\alpha}^{(\textrm{approx})} = \frac{\sum\limits_{m=1}^{m_{\textrm{rec}}^{\textrm{max}}}w_m b_{i\alpha}^{(m)}}{\sum\limits_{m=1}^{m_{\textrm{rec}}^{\textrm{max}}} w_m}
\end{equation}
with $w_m = 1/[\beta(m_{\textrm{rec}}^{\textrm{max}}-m)+1]$ can provide smoother convergence for values of $\beta \ge 1$.

Oscillating values~\cite{Seiser-13} for $a_{i\alpha}^{(m)}$ and $b_{i\alpha}^{(m)}$ can be chosen to treat, e.g., systems with band-gaps~\cite{Turchi-82}.

\subsection{Chebyshev polynomials}
\label{sec:chebyshev}

The Chebyshev polynomials of the second kind in Eq.~\ref{eq:nialpha} are expressed as
\begin{equation}\label{eq:chebyshev}
P_m(\epsilon) = \sum\limits_{n=0}^m p_{mn} \epsilon^n 
\end{equation}
with 
\begin{equation}
p_{(m+1)n} = 2p_{m(n-1)} - p_{(m-1)n}
\end{equation}
(unless $n<0$ or $n>m$ when $p_{mn}=0$). They present the basis of the expansion of $n_{i\alpha}$ (Eq.~\ref{eq:nialpha})~\cite{Drautz-06}.
The values of $P_m(\epsilon)$ are computed iteratively
\begin{equation}\label{eq:chebyshev_iter}
P_{m+1}(\epsilon) = 2\epsilon P_{m}(\epsilon)-P_{m-1}(\epsilon)
\end{equation}
with $P_0=1$ and $P_1=2\epsilon$.
The phase 
\begin{equation}
\epsilon = -\cos \phi 
\end{equation}
transforms the Chebyshev polynomials
\begin{equation}
P_{m}(\epsilon) = \frac{\sin(m+1)\phi}{\sin\phi}
\end{equation}
to sine functions with a corresponding DOS 
\begin{equation}\label{eq:nialpha_sin}
n_{i\alpha}(\epsilon) = \sum\limits_m g_m \sigma_{i\alpha}^{(m)} \sin(m+1) \phi \, .
\end{equation}
This expression can be integrated to provide analytic expressions for the bond energy of orbital $\alpha$ of atom $i$,
\begin{eqnarray}
  U_{{\rm bond}, i\alpha} &=& b_{i\alpha}^{(\infty)} \sum_{m} g_m \sigma_{i\alpha}^{(m)} \left[ \hat{\chi}_{m+2}(\phi_F) \right. \nonumber \\
    & &  \left.- \gamma_0 \hat{\chi}_{m+1}(\phi_F) +\hat{\chi}_m(\phi_F) \right] \, ,
\end{eqnarray}
and the number of electrons
\begin{equation} 
  N_{i\alpha}(\phi_F) = \sum_{m} g_m \sigma_{i\alpha}^{(m)}\hat{\chi}_{m+1}(\phi_F) \, .
\end{equation}
The structure-independent response functions
\begin{equation}
\hat{\chi}_0(\phi_F)=0
\end{equation}
\begin{equation}
 \hat{\chi}_1(\phi_F) =1-\frac{\phi_F}{\pi}+\frac{1}{2\pi}\sin\left( 2\phi_F\right) 
\end{equation}
\begin{equation}
 \label{eq:chi}
 \hat{\chi}_m(\phi_F) =\frac{1}{\pi}\left[ \frac{\sin(m+1)\phi_F}{m+1}-\frac{\sin(m-1)\phi_F}{m-1}\right] 
\end{equation}
with the Fermi phase
\begin{equation}
\label{eq:Fermi-phase}
\cos \phi_F = \frac{E_F - a_{i\alpha}^{(\infty)}}{2 b_{i\alpha}^{(\infty)}}
\end{equation}
correspond to a weighting of the contribution of the structure-dependent expansion coefficients $ \sigma_{i\alpha}^{(m)}$ to the bond energy.

\subsection{Expansion coefficients and moments}
\label{sec:sigma}

The expansion coefficients 
\begin{equation}\label{eq:sigma_i}
  \sigma_{i\alpha}^{(m)} = \sum_{n=0}^{m} p_{mn} \hat{\mu}_{i \alpha}^{(n)}
\end{equation}
in Eq.~\ref{eq:nialpha} carry the information on the atomic structure in the normalised moments ~\cite{Drautz-06}
\begin{equation}
\hat{\mu}_{i \alpha}^{(n)} = \frac{1}{\left(2b_{i\alpha}^{(\infty)}\right)^n}\sum\limits_{l=0}^n 
                             \left(\begin{array}{l} n \\ l\end{array}\right) \left(-1\right)^l {a_{i\alpha}^{(\infty)}}^l \mu_{i\alpha}^{(n-l)}
\end{equation}
with terminator coefficients $a_{i\alpha}^{(\infty)}$ and $b_{i\alpha}^{(\infty)}$ of orbital $\alpha$ of atom $i$.
The moments provide the direct link between the electronic structure, $ n_{i\alpha}(E)$, and the atomic structure by the moments theorem~\cite{CryotLackmann-67}
\begin{eqnarray} \label{eq:momDOS}
\mu^{(n)}_{i\alpha} &=& \int E^{n} n_{i\alpha}(E)dE = \langle i\alpha | \hat{H}^{n} | i\alpha \rangle \\
                       &=& \sum\limits_{j_{1}\beta_{1} ...j_{n-1}\beta_{n-1}} 
                         H_{i\alpha j_{1}\beta_{1}} H_{ j_{1}\beta_{1} j_{2}\beta_{2} }
                         ... H_{ j_{n-1}\beta_{n-1} i\alpha} \nonumber
\end{eqnarray}
This link is schematically illustrated in Fig.~\ref{fig:moments} for the second, third and fourth moment: 
\begin{figure}[htb]
\begin{center}
\includegraphics[width=0.99\columnwidth]{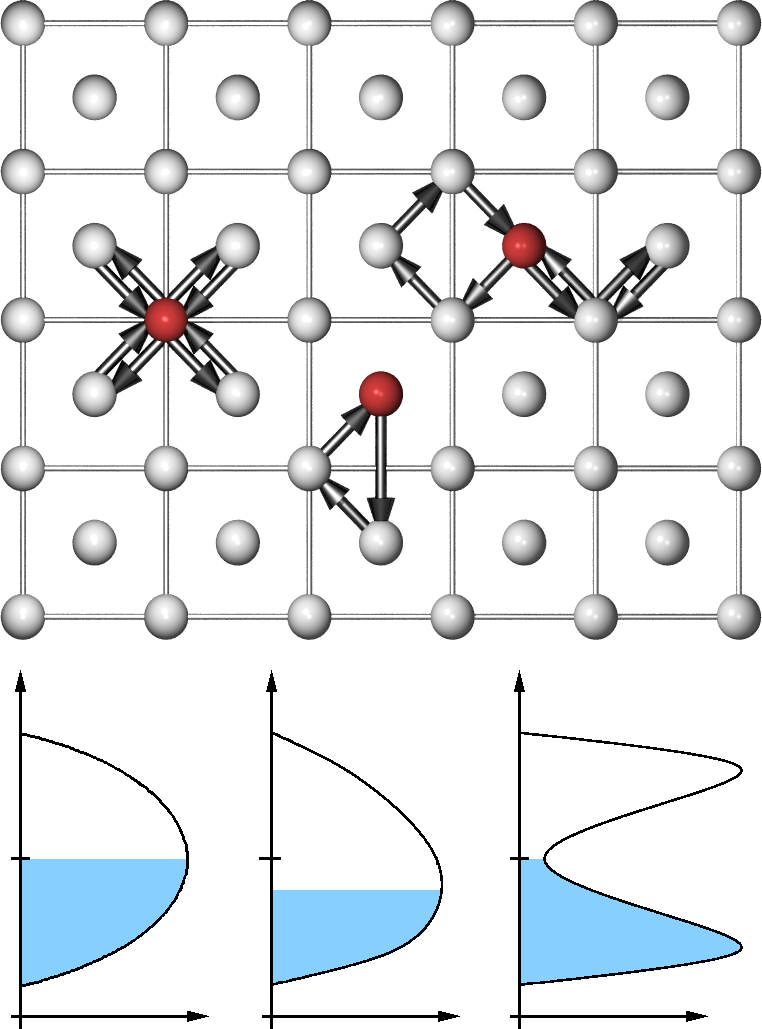}
\caption{\small Schematic illustration of the direct link between the atomic structure in terms of self-returning paths (top) and electronic density of states (bottom).
The second moment that is linked to the RMS width of the DOS (bottom left) is determined by self-returning paths of length two (top, left red atom). The third moment that relates to the skewness of the DOS (bottom middle) is given by paths of length three 
(top, middle red atom) and the fourth moment that is linked to the bimodality of the DOS (bottom right) by paths of length four (top, right red atom).}
\label{fig:moments} 
\end{center}
\end{figure}
The self-returning paths of length two, three and four in the atomic structures are linked to the root mean square (RMS) width, the skewness and the bimodality of the electronic DOS, respectively.

In the intersite representation (Eq.~\ref{eq:TB-Ubond-intersite}), the information on the individual bonds is contained in the bond order $\Theta_{i\alpha j\beta}(\epsilon)$
or the density matrix $n_{i\alpha j\beta}(\epsilon)$ that can be expressed in terms of Chebyshev polynomials~\cite{Drautz-06} 
\begin{eqnarray}\label{eq:nialphajbeta}
\Theta_{i\alpha j\beta}(\epsilon) &=& 2 n_{i\alpha j\beta}(\epsilon) \nonumber \\ 
                                  &=& 2 \frac{2} {\pi} \sqrt{1-\epsilon^{2}} \sum\limits_m g_m \sigma_{i\alpha j\beta}^{(m)} P_{m}(\epsilon)
\end{eqnarray}
with $\sigma_{i\alpha j\beta}^{(m)}$ defined equivalently to Eq.~\ref{eq:sigma_i} as
\begin{equation}
  \sigma_{i\alpha j\beta}^{(m)} = \sum_{n=0}^{m} p_{mn} \hat{\xi}_{i \alpha j \beta}^{(n)}
\end{equation}
and normalization of the interference paths (Eq.~\ref{eq:xi})
\begin{equation}
\hat{\xi}_{i \alpha j \beta}^{(l)} = \frac{1}{\left(2b_{i\alpha}^{(\infty)}\right)^l}\sum\limits_{n=0}^l 
                             \left(\begin{array}{l} n \\ l\end{array}\right) \left(-1\right)^l {a_{i\alpha}^{(\infty)}}^{n-l} \xi_{i\alpha j\beta}^{(n)} \, .
\end{equation}

A \emph{relation between the moments and the atomic structure} is established in the second equality by the self-returning paths 
$i\alpha\rightarrow j_{1}\beta_{1}\rightarrow j_{2}\beta_{2}\rightarrow ... \rightarrow j_{n-1}\beta_{n-1}\rightarrow i\alpha$ 
from orbital $\alpha$ on atom $i$ along orbitals $\beta_k$ of atoms $j_k$ ($k=1 ... n-1$).
Each element of a self-returning path corresponds to the pairwise Hamiltonian matrices in the global coordinate system (Eq.~\ref{eq:Hij_rot}) and carries information about the onsite level of atom $i$
\begin{equation}\label{eq:Hii}
H_{i\alpha i\alpha}=\langle i\alpha|\hat{H}|i\alpha\rangle = E_{i\alpha}
\end{equation}
and the interatomic interactions between the atomic orbitals on neighbouring atoms $i$ and $j$
\begin{equation}
H_{i\alpha j\beta}=\langle i\alpha|\hat{H}|j\beta\rangle \, .
\end{equation}
Higher moments correspond to longer paths and thus to a more far-sighted sampling of the atomic environment.
As different crystal structures have different sets of self-returning paths of a given length, the moments 
may be seen as fingerprints of the crystal structure~\cite{Hammerschmidt-16,Turchi-91} and used to construct maps of structural similarity~\cite{Jenke-18}. 

The paths can be computed efficiently by realizing that (1) only the sum of all paths is relevant (Eq.~\ref{eq:momDOS}) and that (2) 
the sums across the whole paths can be represented as sums along path segments. 
The path segments are the interference paths
\begin{equation} 
\label{eq:xi}
\xi_{i\alpha j\beta}^{(n)} = \langle i\alpha | \hat{H}^{n} |  j\beta \rangle \,
\end{equation}
of length $n$ between atom $i$ and $j$. 
The computation of interference paths can be simplified after realising that they can be (i) constructed iteratively
\begin{equation}\label{eq:xi_rec}
\xi_{i\alpha j\beta}^{(n)} = \sum_{k\gamma} H_{i\alpha k\gamma}  \xi^{(n-1)}_{k\gamma j\beta} 
\end{equation}
for all interaction neighbours $k$ with orbitals $\gamma$,
(ii) inverted in their direction by taking the transpose
\begin{equation}\label{eq:xi_invert}
\xi_{i\alpha j\beta}^{(n)} = {\xi_{j\beta i\alpha}^{(n)}}^T
\end{equation}
and (iii) merged by multiplication of segments
\begin{equation}\label{eq:xi_merge}
\xi_{i\alpha j\beta}^{(n)} = \sum_{k\gamma} \xi^{(l)}_{i\alpha k\gamma}  \xi^{(n-l)}_{k\gamma j\beta} 
\end{equation}
of length $0<l<n$ for all common endpoint atoms $k$ with orbital $\gamma$.
Using these properties, the summation of matrix multiplications along the individual self-returning paths can
be decomposed to segments that represent summations of matrix multiplications along shorter partial paths. 
It is, therefore, not necessary to determine each possible path $\xi_{i\alpha j\beta}^{(n)}$ between atoms $i$ and $j$ individually, 
but instead sufficient to determine the set of shorter segments that is needed for their construction.
The implementation of this approach in BOPfox reaches the theoretical scaling limits of the required execution time 
and is discussed in detail in Ref.~\cite{Teijeiro-16-1}.

A \emph{relation between the moments and the electronic structure} is due to the expansion coefficients $a_{i\alpha}^{(n)}$ and $b_{i\alpha}^{(n)}$ ~\cite{Aoki-93-2}.
These coefficients determine the electronic structure in terms of $n_{i\alpha}$, the local DOS, as given in Eq.~\ref{eq:DOS-G}.
The first four moments of the local DOS are given by 
\begin{eqnarray}
\mu_{i\alpha}^{(0)} &=& 1 \\
\mu_{i\alpha}^{(1)} &=& a_{i\alpha}^{(0)}  \\
\mu_{i\alpha}^{(2)} &=& {a_{i\alpha}^{(0)}}^2 + {b_{i\alpha}^{(1)}}^2  \\
\mu_{i\alpha}^{(3)} &=& {a_{i\alpha}^{(0)}}^3 + 2a_{i\alpha}^{(0)} {b_{i\alpha}^{(1)}}^2 + a_{i\alpha}^{(1)} {b_{i\alpha}^{(1)}}^2 
\end{eqnarray}
which is easily verified by identifying all self-returning paths of corresponding length in Fig.~\ref{fig:linear-chain}.
Vice-versa, the recursion coefficients can be determined from the moments~\cite{Haydock-80-2,Horsfield-96-3} for each $i\alpha$ by
\begin{equation}
  a_n = \sum_{j=0}^n \sum_{l=0}^n c_j^n c_l^n \mu^{j+l+1}
  \label{eq:momtoa}
\end{equation}
and
\begin{equation}
  b_n = \sum_{j=0}^n \sum_{l=0}^{n-1} c_j^n c_l^{n-1} \mu^{j+l+1}
  \label{eq:momtob}
\end{equation}
where we dropped the common index $i\alpha$ for readability. The coefficients $c_j^n$ are given by
\begin{equation*}
  c_0^0 = 1,
\end{equation*}
\begin{equation*}
  c_j^n = 0  \text{\quad if } j>n \text{ or } j<0 \text{ or } n<0,
\end{equation*}
\begin{equation*}
 b_{n+1}c_j^{n+1} = c_{j-1}^{n} - a_n c_j^n - b_n c_j^{n-1}
 \label{eq:recursion_bc}
\end{equation*}
and determined iteratively.

\subsection{Damping factors}
\label{sec:g_m}

The damping factors $g_m$ in Eq.~\ref{eq:nialpha}, together with approximate higher expansion coefficients,
were introduced to suppress Gibbs ringing and ensure strictly positive values of the DOS~\cite{Seiser-13}.
Therefore the calculation of the DOS (Eq.~\ref{eq:nialpha}) with expansion coefficients $\sigma_{i\alpha}^{(m)}$ 
from the moments up to $m=n_{\rm max}$ (Eq.~\ref{eq:sigma_i}) is expanded up to $n_{\rm max}+1<m<n_{\rm exp}$ 
with estimated higher expansion coefficients $\sigma_{i\alpha}^{(m)}$~\cite{Seiser-13}
\begin{eqnarray}
n_{i\alpha}^{(n_{\rm max})}(\epsilon) = &\frac{2} {\pi} \sqrt{1-\epsilon^{2}}  
              \left[ \sum\limits_{m=1}^{n_{\rm max}} g_m \sigma_{i\alpha}^{(m)} P_{m}(\epsilon) \right. \nonumber \\
            &            \left. +\sum\limits_{m=n_{\rm max}+1}^{n_{\rm exp}} g_m \sigma_{i\alpha}^{(m)} P_{m}(\epsilon) \right] \, .
\end{eqnarray}
The higher expansion coefficients $\sigma_{i\alpha}^{(m)}$ are obtained by recursive calculation of the interference paths~\cite{Seiser-13}
along the semi-infinite chain,
\begin{equation}
\zeta_k^{(m+1)} = 2 \left[ \hat{a}_k \zeta_k^{(m)} + \hat{b}_k \zeta_{k-1}^{(m)} + \hat{b}_{k+1} \zeta_{k+1}^{(m)} \right] - \zeta_k^{(m-1)}
\end{equation}
with 
\begin{equation}
\label{eq:aexp}
\hat{a}_k=\frac{a_k-a_{i\alpha}^{(\infty)}}{2b_{i\alpha}^{(\infty)}} \quad {\rm and} \quad
\hat{b}_k=\frac{b_k}{2b_{i\alpha}^{(\infty)}} 
\end{equation}
and $\sigma_{i\alpha}^{(n)}=\zeta_{0}^{(n)}$.
The relative importance of the higher, approximated expansion coefficients with respect to the lower, computed ones is
balanced by the damping factors $g_m$ in Eq.~\ref{eq:nialpha} that vary smoothly from 1 to 0.
In BOPfox, the Jackson kernel~\cite{Weisse-06} 
\begin{equation}
\label{eq:Jackson}
g^J_m = \frac
        { \left(n_{\mathrm{max}}-m+1\right) \cos\frac{\pi m}{n_{\mathrm{max}}+1} + \sin \frac{\pi m}{n_{\mathrm{max}}+1}\cot\frac{\pi} {n_{\mathrm{max}}+1} }
        {n_{\mathrm{max}}+1}
\end{equation}
for an expansion $m=1\dots n_{\rm max}$ is adapted to Chebyshev polynomials of the second kind by
\begin{equation}
g_m = g^J_{m+1}/g^J_{1}
\end{equation}
as described in Ref.~\cite{Seiser-13}.

\subsection{Band-width estimates}
\label{sec:terminator}

The different terminators (Eqs.~\ref{eq:abinf} and~\ref{eq:weighted}) of the continued fraction (Eq.~\ref{eq:DOS-G}) 
require the recursion coefficients $a_{i\alpha}^{(m)}$ and $b_{i\alpha}^{(m)}$ beyond the ones that can be computed 
from the $\mu_{i\alpha}^{(n)}$ with $m>n$ by Eqs.~\ref{eq:momtoa} and~\ref{eq:momtob}.
We determine approximate values of $a_{i\alpha}^{(m)}$ and $b_{i\alpha}^{(m)}$ from estimates of the centre and the width of the DOS
\begin{eqnarray}
a_{i\alpha}^{(\infty)} = & A_{i\alpha}^{(\infty)} \\
b_{i\alpha}^{(\infty)} = & B_{i\alpha}^{(\infty)}
\end{eqnarray}
with
\begin{eqnarray}
A_{i\alpha}^{(\infty)} =& \frac{1}{2} (E_{i\alpha}^{\rm top} + E_{i\alpha}^{\rm bottom}) \\
B_{i\alpha}^{(\infty)} =& \frac{1}{4} (E_{i\alpha}^{\rm top} - E_{i\alpha}^{\rm bottom}) \, .
\end{eqnarray}
The values of $A_{i\alpha}^{(\infty)}$ and $B_{i\alpha}^{(\infty)}$ can be estimated in BOPfox in several ways 
based on the computed recursion coefficients $a_{i\alpha}^{(n)}$ and $b_{i\alpha}^{(n)}$ for $n$ levels of orbital $\alpha$ on atom $i$.
The simple approximations are (i) the lowest computed recursion coefficients, i.e., 
\begin{equation}
A_{i\alpha}^{(\infty)} = a_{i\alpha}^{(1)} \quad , \quad B_{i\alpha}^{(\infty)}= b_{i\alpha}^{(1)} \, ,
\end{equation}
(ii) the highest computed recursion coefficients
\begin{equation}
A_{i\alpha}^{(\infty)} = a_{i\alpha}^{(n_{\rm max})} \quad , \quad B_{i\alpha}^{(\infty)}= b_{i\alpha}^{(n_{\rm max})} \, ,
\end{equation}
(iii) averaged values~\cite{Ford-14} similar to Haydock and Johannes~\cite{Haydock-75-1},
\begin{equation}
A_{i\alpha}^{(\infty)} = \frac{\sum\limits_{n=0}^{n_{\rm max}} a_{i\alpha}^{(n)}}{{n_{\rm max}} +1 } \quad , \quad 
B_{i\alpha}^{(\infty)} = \sqrt{\frac{\sum\limits_{n=1}^{n_{\rm max}} {b_{i\alpha}^{(n)}}^2}{n_{\rm max}}} \, ,
\end{equation}
(iv) the average band-centre with the band-width from the highest computed recursion level
\begin{equation}
A_{i\alpha}^{(\infty)} = \frac{\sum\limits_{n=0}^{n_{\rm max}} a_{i\alpha}^{(n)}}{{n_{\rm max}} +1 } \quad , \quad
B_{i\alpha}^{(\infty)} = b_{i\alpha}^{(n_{max})} \, ,
\end{equation}
or (v) lowest computed band-bottom and highest computed band-top~\cite{Ford-14}
\begin{eqnarray}
\label{eq:findeminemax}
A_{i\alpha}^{(\infty)} =& \frac{ {\rm max} \left( a_{i\alpha}^{(n)} \right) + {\rm min} \left( a_{i\alpha}^{(n)} \right) }  {2} \quad , \\
B_{i\alpha}^{(\infty)} =& \frac{ {\rm max} \left( a_{i\alpha}^{(n)} \right) - {\rm min} \left( a_{i\alpha}^{(n)} \right)  + 4 {\rm max} \left( b_{i\alpha}^{(n)} \right) }{4} \, . 
\end{eqnarray}
Further choices are (vi) the approach of Beer {\it et al.}~\cite{Beer-84} 
that minimises the band-width with preserved moments of the DOS
and (vii) Gershogorin's circle theorem~\cite{Gerschogorin-31} which leads to estimates of the band-edges~\cite{Seiser-13}
\begin{eqnarray}
\label{eq:gershogorin}
E_{i\alpha}^{\rm bottom} &=& {\rm min}\left( a_{i\alpha}^{(n)}-b_{i\alpha}^{(n)}-b_{i\alpha}^{(n+1)} \right) \\
E_{i\alpha}^{\rm top}    &=& {\rm max}\left( a_{i\alpha}^{(n)}+b_{i\alpha}^{(n)}+b_{i\alpha}^{(n+1)} \right) \, .
\end{eqnarray}
For testing purposes the user can also define (viii) global values of $A^{(\infty)}$ and $B^{(\infty)}$ that hold for all atoms.

\subsection{Example with typical settings: bcc Ta}

\begin{figure}[htb]
\begin{center}
\includegraphics[width=0.99\columnwidth]{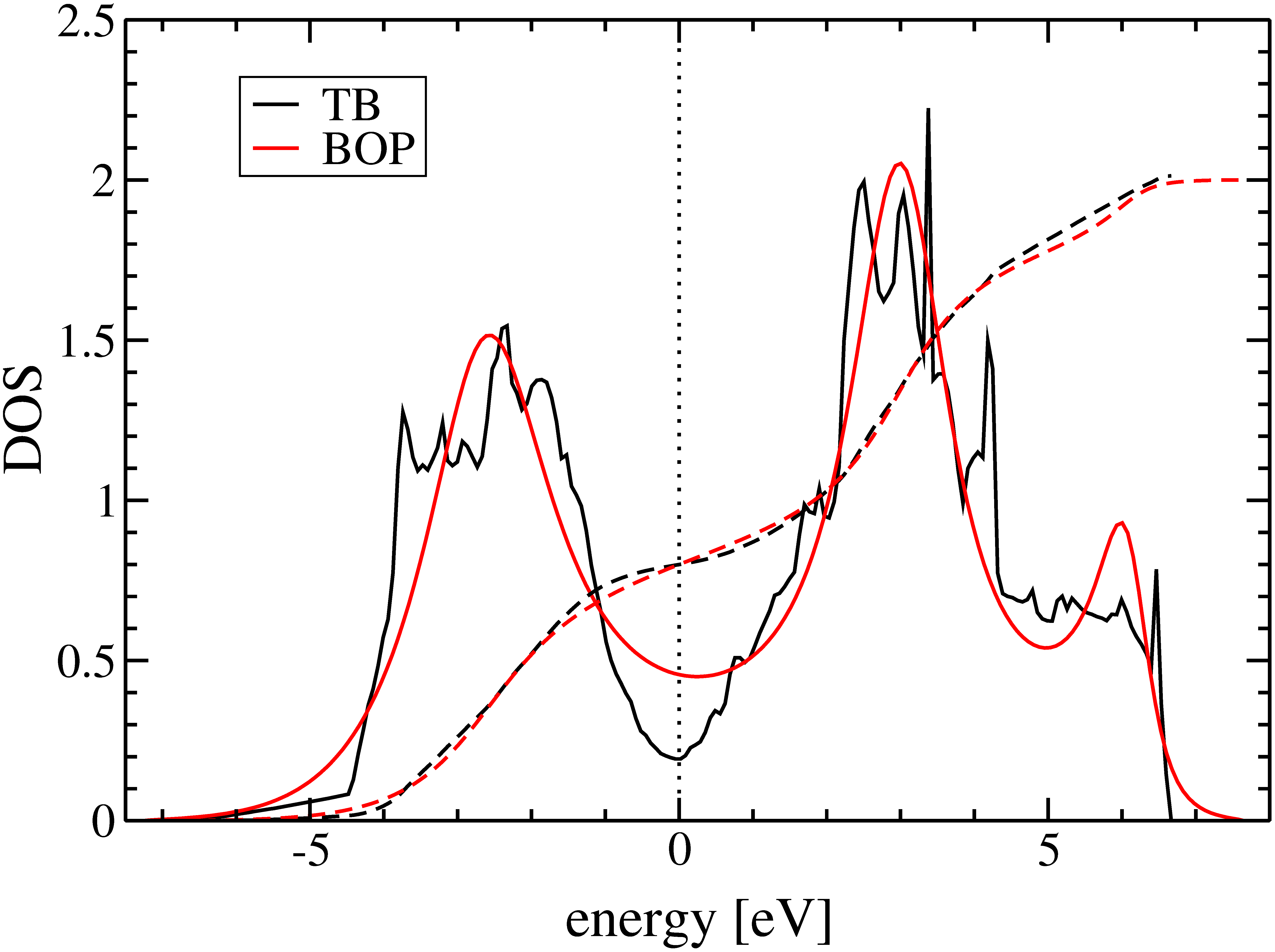}
\caption{\small DOS of bcc Ta computed by reciprocal-space TB (black) and real-space BOP (red) computed with the parametrisation of Ref.~\cite{Cak-14}.
The integrated DOS of TB and BOP (divided by a factor of five for plotting convenience) are given as dashed lines. The dotted line marks the Fermi level.}
\label{fig:DOS} 
\end{center}
\end{figure}

As an example of an analytic BOP calculation, we used the parametrisation of Ref.~\cite{Cak-14} to determine the 
DOS of bcc Ta shown in Fig.~\ref{fig:DOS}.
This non-magnetic BOP calculation with a $d$-band model uses 9 moments (Eq.~\ref{eq:momDOS}), a square-root terminator (Eq.~\ref{eq:abinf}),
the Gershogorin bandwidth estimate (Eq.~\ref{eq:gershogorin}), and estimated expansion coefficients up to moment 200 (Eq.~\ref{eq:aexp})
that are damped with a Jackson kernel (Eq.~\ref{eq:Jackson}). 
The DOS obtained by analytic BOPs is in good agreement with the TB reference (20$\times$20$\times$20 ${\mathbf k}$-point mesh, tetrahedron integration). 
In both cases, the Fermi level is in the pseudo-gap of the bimodal DOS that is typical for bcc transition metals.
The bandwidth of the DOS, as well as the position and height of the two most prominent peaks are well captured.
The integrated DOS of analytic BOPs is in excellent agreement with the TB reference which is the basis for reproducing DOS-integral quantities like the bond energy.

\section{Forces and torques in analytic BOPs}
\label{sec:BOP_forces}
\setcounter{figure}{0}    

\subsection{General binding-energy derivative}

The minimisation of the binding energy in the self-consistency cycle (see~\ref{sec:SCF}) is based on the derivative of the binding energy with respect to onsite-levels $E_{j\beta\mu}$.
The computation of forces and stresses requires the derivative of the binding energy with respect to the position $\mathbf{r}_j$,
while determining the torques makes use of the derivative of the binding energy with respect to the spin orientation $\mathbf{s}_{j\beta\mu}$ of atom $j$.
These are all specific examples of derivatives of the binding energy which can be written in a generic form as derivatives with respect to a general parameter $\Lambda$,~\cite{Drautz-11}
\begin{eqnarray}
\label{eq:dUdLambda}
\hspace{-0.5cm}
\frac{dU_B}{d\Lambda}&=&  \sum\limits_{i\alpha\nu} \sum\limits_{n=0}^{n_{\rm max}} w^{(n)}_{i\alpha\nu} \frac{d\mu_{i\alpha\nu}^{(n)}}{d\Lambda}  \nonumber \\
                      &&   - \sum\limits_{i\alpha\nu} N_{i\alpha\nu} \frac{dE_{i\alpha\nu}}{d\Lambda}  + \frac{d U_{\rm rep}}{d\Lambda} \, .
\end{eqnarray}
The total derivative of the bond energy $U_{\rm bond}$ (Eq.~\ref{eq:TB-Ubond-intersite}) with respect to $\Lambda$ is transformed to
partial derivatives with respect to moments $\mu_{i\alpha\nu}^{(n)}$ and associated partial derivatives of the moments with respect to $\Lambda$.
This allows the derivative of the bond energy to be expressed in the form of Hellmann-Feynman-type forces
\begin{equation}
 \frac{dU_{\rm bond}}{d\Lambda} = \sum\limits_{i\alpha\nu j\beta\mu} {\tilde \Theta}_{i\alpha\nu j\beta\mu} \frac{dH_{j\beta\mu i\alpha\nu}}{d\Lambda}
\end{equation}
with a bond-order-like term
\begin{equation}\label{eq:theta-tilde}
  {\tilde \Theta}_{i\alpha\nu j\beta\mu} = \sum\limits_{n=1}^{n_{\rm max}} \Xi_{i\alpha\nu j\beta\mu}^{(n-1,n)} \, .
\end{equation}
Inserting $E_{j\beta\mu}$ for $\Lambda$ leads to the self-consistency condition of Eq.~\ref{eq:BOP_SCF}.
Replacing $\Lambda$ with $\mathbf{r}_j$ or $\mathbf{s}_{j\beta}$ yields forces and torques as described in~\ref{sec:forces} and~\ref{sec:torques}, respectively. 

The derivatives of $U_{\rm bond}$ with respect to the moments 
\begin{equation}
\label{eq:weights}
w^{(n)}_{i \alpha\nu} = \frac{\partial U_{bond}}{\partial {\mu}_{i \alpha\nu}^{(n)} }
\end{equation} 
enter ${\tilde \Theta}_{i\alpha\nu j\beta\mu}$ as weights $w_{i\alpha\nu}^{(m)}$ in
\begin{eqnarray}
\Xi_{i_1 \alpha_1\nu_1 i_n \alpha_n\nu_n}^{(n-1,m)} =& \sum_{ i_2\alpha_2\nu_2 \dots i_{n-1}\alpha_{n-1}\nu_{n-1}}
                                                         \left( \sum_{l=1}^n w_{i_l\alpha_l\nu_l}^{(m)} \right) \nonumber \\
                                                     &  H_{i_1\alpha_1\nu_1 i_2\alpha_2\nu_2} \dots H_{i_{n-1}\alpha_{n-1}\nu_{n-1} i_n\alpha_n\nu_n}
\end{eqnarray}
and are given in detail in~\ref{sec:weights}.
This compact form leads to an efficient recursive computation of ${\tilde \Theta}_{i\alpha j\beta}$ by
\begin{equation}\label{eq:Xi}
\Xi_{i \alpha j \beta}^{(n-1,m)} = T_{i \alpha j \beta}^{(n-1,m)} + w_{i \alpha}^{(m)} \xi_{i \alpha j \beta}^{(n-1)}
\end{equation}
with transfer paths $T_{i \alpha j \beta}^{(n,m)}$. 
The transfer paths are closely related to the interference paths (Eq.~\ref{eq:momDOS})
and exhibit similar properties (Eqs.~\ref{eq:xi_rec}-\ref{eq:xi_merge}).
In particular, the transfer paths can also be (i) constructed iteratively
\begin{equation}
T_{i\alpha j\beta}^{(n,m)} = \sum_{k \gamma} H_{i\alpha k\gamma} T_{k\gamma j\beta}^{(n-1,m)}  
                               +  w_{i\alpha}^{(m)} \xi_{i\alpha j\beta}^{(n)} \, ,
\end{equation}
(ii) inverted by taking the transpose
\begin{equation}
T_{i\alpha j\beta}^{(n,m)} = {T_{j\beta i\alpha}^{(n,m)}}^T \, ,
\end{equation}
and (iii) merged by a product rule
\begin{equation}\label{eq:Xi-merge}
T_{i\alpha j\beta}^{(n-1,m)} = \sum\limits_{k\gamma} T_{i\alpha k\gamma}^{(l-1,m)}\xi_{k\gamma j\beta}^{(n-l)}
                             + \sum\limits_{k\gamma} \xi_{i\alpha k\gamma}^{(l-1)} T_{k \gamma j \beta}^{(n-l,m)}  \, .
\end{equation}
These properties of the transfer paths are the basis for the efficient~\cite{Teijeiro-16-1} and parallel~\cite{Teijeiro-16-2,Teijeiro-17}
implementation of self-consistency, forces and torques in analytic BOPs. 

For \emph{non-collinear magnetism}, the above equations are transformed by rewriting the moments and weights as 2$\times$2 matrices 
in spin space (see~\ref{sec:magnetism}). The general derivative of the binding energy (Eq.~\ref{eq:dUdLambda}) reads~\cite{Ford-15}
\begin{eqnarray}
\label{eq:dUdLambda-noncollinear}
\hspace{-0.5cm}
\frac{dU_B}{d\Lambda}&=&  \sum\limits_{i\alpha} \sum\limits_{n=0}^{n_{\rm max}} {\rm Tr} \left( {\bm w}^{(n)}_{i\alpha} \frac{d{\bm \mu}_{i\alpha}^{(n)}}{d\Lambda}\right)  \nonumber \\
                      &&   - \sum\limits_{i\alpha\nu} N_{i\alpha\nu} \frac{dE_{i\alpha\nu}}{d\Lambda}  + \frac{d U_{\rm rep}}{d\Lambda}
\end{eqnarray}
with
\begin{equation}
  \frac{d{\bm \mu}_{i\alpha}^{(n)}}{d\Lambda} = 
    \left( \begin{array}{cc}
    \frac{d\mu_{i\alpha}^{\uparrow\uparrow(n)}}{d\Lambda}    & \frac{d\mu_{i\alpha}^{\uparrow\downarrow(n)}}{d\Lambda}  \\
    \frac{d\mu_{i\alpha}^{\downarrow\uparrow(n)}}{d\Lambda}  & \frac{d\mu_{i\alpha}^{\downarrow\downarrow(n)}}{d\Lambda}   
  \end{array} 
  \right)
\end{equation}
and weights that are constructed in the local frame
\begin{equation}
  {\bm w}_{i\alpha}^{(n,{\rm local})} 
  = {\bm U}_{i\alpha}^{ } {\bm w}_{i\alpha}^{(n)} {\bm U}_{i\alpha}^\dagger
  = \left( \begin{array}{cc}
  w_{i\alpha}^{\uparrow(n,{\rm local})} & 0 \\
  0                           & w_{i\alpha}^{\downarrow(n,{\rm local})} 
  \end{array}
  \right)
\end{equation}
from the global counterparts by a unitary transformation like the onsite levels (Eq.~\ref{eq:onsite-noncollinear}).
The transformation of the bond order term ${\tilde \Theta}_{i\alpha\nu j\beta\mu}$ and the transfer matrices 
$T_{i\alpha j\beta}^{(n-1,m)}$ to 2$\times$2 spin space leads to the same equations as Eq.~\ref{eq:theta-tilde}
and Eq.~\ref{eq:Xi}, respectively with corresponding interference paths
\begin{equation}
   {\bm \xi}_{i\alpha j\beta}^{(n)} = 
    \left( \begin{array}{cc}
    \xi_{i\alpha j\beta}^{\uparrow\uparrow(n)} & \xi_{i\alpha j\beta}^{\uparrow\downarrow(n)} \\
    \xi_{i\alpha j\beta}^{\downarrow\uparrow(n)} & \xi_{i\alpha j\beta}^{\downarrow\downarrow(n)}
  \end{array} 
  \right) \, .
\end{equation}

\subsection{Forces}
\label{sec:forces}

Replacing the derivative $d/d\Lambda$ in Eq.~\ref{eq:dUdLambda} with the gradient $\nabla_k$ leads to the analytic forces. 
With self-consistent onsite levels $dU/dE_{i\alpha}=0$ (Eq.~\ref{eq:BOP_SCF}), 
the forces on atom $k$ in TB and BOP calculations are given by~\cite{Drautz-11,Ford-14,Ford-15}
\begin{eqnarray} \label{eq:F_k}
  {\bm F}_k &=& -\nabla_k U_B  \nonumber \\
            &=& - \sum\limits_{i\alpha j\beta}^{i\alpha \ne j\beta} {\tilde \Theta}_{i\alpha j\beta} \nabla_k H_{j\beta i\alpha} \nonumber \\
            & & -\frac{1}{2} \sum\limits_{i\alpha j\beta} (\nabla_k J_{i\alpha j\beta}) q_{j\beta} q_{i\alpha} \nonumber \\
            & & +\frac{1}{4} \sum\limits_{i\alpha j\beta} (\nabla_k I_{i\alpha j\beta}) \mathrm{m}_{j\beta} \mathrm{m}_{i\alpha} \nonumber \\
            & & - \nabla_k U_{\rm rep} \, .
\end{eqnarray}
(This expression also holds for non-collinear magnetism as there only the onsite levels are affected by the rotation~\cite{Ford-15}.)
In TB calculations this expression corresponds to Hellmann-Feynman forces~\cite{Hellmann-37,Feynman-39} and
\begin{equation}
 \label{eq:Theta_tilde_BOP}
 {\tilde \Theta}_{i\alpha j\beta} = \frac{\partial U_B}{\partial H_{i\alpha j\beta}} \,.
\end{equation}
becomes the density matrix $n_{i\alpha j\beta}$ (Eq.~\ref{eq:TB-Ubond-intersite}). 
In analytic BOP calculations, in contrast, the approximate evaluation of the DOS means that a self-consistent set of charges and magnetic 
moments does not correspond to a stationary point in the BOP energy (as it does in DFT or TB approaches)~\cite{Drautz-11}.
However, taking exact derivatives of the energy with respect to atomic positions, 
this form can still be used to represent forces~\cite{Drautz-11,Ford-14} (Eq.~\ref{eq:F_k}) and stresses~\cite{Schreiber-inprep}.
The contribution of the bond energy to the atomic virial stress is~\cite{Schreiber-inprep}
\begin{equation} \label{eq:sigma_k}
  {\bm \sigma}^{(i)}_{\rm bond} = \frac{1}{2} \sum\limits_{\alpha j \beta} {\tilde \Theta}_{i\alpha j\beta} 
                                  \left( \nabla_j H_{j\beta i\alpha} - \nabla_i H_{j\beta i\alpha} \right) \otimes r_{ij} \, .
\end{equation}
%

\subsection{Torques}
\label{sec:torques}

Inserting the local spin direction ${\bm s}_{i\alpha}$ for $\Lambda$ in the general derivative (Eq.~\ref{eq:dUdLambda}) leads
to the torques, i.e., to the change in binding energy due to rotation of local spin directions. 
The derivatives of the rotation matrices can be taken into account by expressing the weights in terms of their local counterparts~\cite{Ford-15}
\begin{equation}
{\bm w}_{i\alpha}^{(n)} = \left[ 
                          \frac{1}{2} \left( w_{i\alpha}^{\uparrow}   + w_{i\alpha}^{\downarrow} \right) {\bm 1} 
                          + \frac{1}{2} \left( w_{i\alpha}^{\downarrow} - w_{i\alpha}^{\uparrow}   \right) {\bm s}_{i\alpha} \cdot {\bm \sigma}
                          \right]
\end{equation}
where we dropped the index $(n,{\rm local})$ of $w_{i\alpha}$ for brevity.
With $\Delta_{i\alpha}^{ } = E_{i\alpha}^{\downarrow} - E_{i\alpha}^{\uparrow}$, the derivative of the bond energy with respect to ${\bm s}_{i\alpha}$
is given by~\cite{Ford-15} 
\begin{eqnarray}
\frac{{\rm d} U_B}{{\rm d}{\bm s}_{i\alpha}} =& \frac{1}{2} \left( {\rm Tr} \left( {\bm {\tilde \Theta}}_{i\alpha i\alpha}  {\bm \sigma} \right) \Delta_{i\alpha} 
                                               -I_i m_i {\bm m}_i \right. \nonumber \\
                                              & \left. + \sum\limits_n^{n_{\rm max}} \left( w_{i\alpha}^{\downarrow} - w_{i\alpha}^{\uparrow} \right) 
                                                {\rm Tr} \left( {\bm \sigma}\mu_{i\alpha}^{(n)} \right) \right)
\end{eqnarray}
where ${\bm m}_i$ is the spin direction on atom $i$ and ${\bm \sigma}$ is the vector of Pauli spin matrices.
The cross product with the spin direction leads to the magnetic torque~\cite{Gilbert-04} given by 
\begin{equation}\label{eq:torque}
{\bm t}_{i\alpha} = \frac{{\rm d} U_B}{{\rm d}{\bm s}_{i\alpha}} \times {\bm s}_{i\alpha}
\end{equation}
for orbital $\alpha$ on atom $i$ where ${\bm s}_{i\alpha} \times {\bm m}_{i\alpha} = 0$.

\subsection{Common partial derivatives}
\label{sec:weights}

The weights (Eq.~\ref{eq:weights}) can be determined analytically.
To this end, the band and onsite contributions are separated as~\cite{Ford-14}
\begin{equation}
\begin{aligned}
w_{i\alpha\nu}^{(n)}= & \frac{\partial U_{\rm band}}{\partial \mu^{(n)}_{i\alpha\nu}} + 
                         \frac{\partial}{\partial \mu_{i\alpha\nu}^{(n)}} \int\limits^{E_F} n_{i\alpha\nu} \mathrm{d}E 
                         \cdot \biggl( E_{i\alpha\nu} \nonumber \\
                      & - \frac{\sum_{j\beta\mu} E_{j\beta\mu} n_{j\beta\mu}(E_F)}{\sum_{j\beta\mu} n_{j\beta\mu}(E_F)} J_iq_i 
                          - \frac{\sum_{j\beta} J_jq_j n_{j\beta}(E_F)}{\sum_{j\beta} n_{j\beta}(E_F)}  \\
                      & -\frac{1}{2}\biggl( (-1)^{\nu} I_i m_i  
                         -\frac{\sum_{j\beta\mu} (-1)^\mu I_j m_j n_{j\beta\mu} (E_F)}{\sum_{j\beta\mu} n_{j\beta\mu} (E_F)}\biggr)\biggr) \\
\end{aligned}
\end{equation}
with 
\begin{equation}
\begin{aligned}
\frac{\partial U_{\rm band}}{\partial \mu^{(n)}} & =  \sum\limits_{m=0}^{n_{\mathrm max}} 
                      \biggl(\frac{\partial b^{(\infty)}}{\partial \mu^{(n)}} \sigma^{(m)}
                             \bigl[\hat{\chi}_{m+2} - 2\epsilon_{F}\hat{\chi}_{m+1} + \hat{\chi}_m \bigr] \\
  & +b^{(\infty)} \biggl( \frac{\partial \sigma^{(n)}}{\partial \mu^{(n)}}
                                     + \frac{\partial \sigma^{(n)}}{\partial a^{(\infty)}}
                                       \frac{\partial a^{(\infty)}}{\partial \mu^{(n)}}
                                     + \frac{\partial \sigma^{(n)}}{\partial b^{(\infty)}}
                                       \frac{\partial b^{(\infty)}}{\partial \mu^{(n)}} \biggr) \\
  & \cdot \bigl[\hat{\chi}_{m+2} - 2\epsilon_{F}\hat{\chi}_{m+1} + \hat{\chi}_m \bigr] \\
  & + \biggl[ \frac{\partial \hat{\chi}_{m+2}}{\partial a^{(\infty)}} 
      - 2 \frac{\partial \epsilon_F}{\partial a^{(\infty)}}\hat{\chi}_{m+1}
      - 2 \epsilon_F \frac{\partial \hat{\chi}_{m+1}}{\partial a^{(\infty)}}
      + \frac{\partial \hat{\chi}_m}{\partial a^{(\infty)}} \biggr] \\
  & \cdot b^{(\infty)}\sigma^{(m)} \frac{\partial a^{(\infty)}}{\partial \mu^{(n)}} \\
  & + \biggl[ \frac{\partial \hat{\chi}_{m+2}}{\partial b^{(\infty)}} 
      - 2 \frac{\partial \epsilon_F}{\partial b^{(\infty)}}\hat{\chi}_{m+1}
      - 2 \epsilon_F \frac{\partial \hat{\chi}_{m+1}}{\partial b^{(\infty)}}
      + \frac{\partial \hat{\chi}_m}{\partial b^{(\infty)}} \biggr] \\
  & \cdot b^{(\infty)}\sigma^{(m)} \frac{\partial b^{(\infty)}}{\partial \mu^{(n)}} \biggr)
\end{aligned}
\end{equation}
where a constant terminator (Eq.~\ref{eq:abinf}) was assumed and 
\begin{equation}
\begin{aligned}
\frac{\partial}{\partial \mu^{(n)}} & \int\limits^{E_F} n(E)dE =  \sum\limits_{m=0}^{n_{\mathrm max}} \biggl( \hat{\chi}_{m+1}(\phi_F) \\
  & \cdot \biggl(   \frac{\partial \sigma^{(m)}}{\partial \mu^{(n)}} 
                  + \frac{\partial \sigma^{(m)}}{\partial a^{(\infty)}} \frac{\partial a^{(\infty)}}{\partial \mu^{(n)}}
                  + \frac{\partial \sigma^{(m)}}{\partial b^{(\infty)}} \frac{\partial b^{(\infty)}}{\partial \mu^{(n)}} \biggr) \\
  & + \sigma^{(m)} \biggl( 
                    \frac{\partial \hat{\chi}_{m+1}}{\partial a^{(\infty)}} \frac{\partial a^{(\infty)}}{\partial \mu^{(n)}}
                  + \frac{\partial \hat{\chi}_{m+1}}{\partial b^{(\infty)}} \frac{\partial b^{(\infty)}}{\partial \mu^{(n)}} \biggr) \biggr) \, .
\end{aligned}
\end{equation}
where we omitted the common index $i\alpha\nu$ for brevity. 
The partial derivatives of the expansion coefficients with respect to the moments are given by
\begin{equation}
\frac{\partial \sigma_{ }^{(m)}} {\partial \mu_{ }^{(n)}}  
  = \sum\limits_{k=n}^{m} p_{mk} \frac{\partial \hat{\mu}_{ }^{(k)}}{\partial \mu_{ }^{(n)}} 
  = \sum\limits_{k=n}^{m} \frac{p_{mk}}{{\left( 2b_{ }^{(\infty)}\right)}^k} (kn)\left( -a_{ }^{(\infty)}\right)^{(k-n)}
\end{equation}
and with respect to the asymptotic recursion coefficients
\begin{eqnarray}
\frac{\partial \sigma_{ }^{(m)}} {\partial a_{ }^{(\infty)}} 
  &=& \sum\limits_{k=0}^{m} p_{mk} \frac{\partial \hat{\mu}_{ }^{(k)}}{\partial a_{ }^{(\infty)}} \nonumber \\
  &=& - \sum\limits_{k=n}^{m} \frac{p_{mk}}{{\left( 2b_{ }^{(\infty)}\right)}^k} \nonumber \\
  & &   \sum\limits_{n=0}^{k-1} (k-n) \left( \begin{array}{c} k \\ n \end{array} \right) \mu_{ }^{(n)} \left( -a_{ }^{(\infty)}\right)^{(k-n-1)}
\end{eqnarray}
\begin{eqnarray}
\label{eq:dsigma_db}
\frac{\partial \sigma_{ }^{(m)}} {\partial b_{ }^{(\infty)}}  
  &=& \sum\limits_{k=0}^{m} p_{mk} \frac{\partial \hat{\mu}_{ }^{(k)}}{\partial b_{ }^{(\infty)}}
      = - \sum\limits_{k=1}^{m} k \frac{p_{mk}} {b_{ }^{(\infty)}} \hat{\mu}_{ }^{(k)}
\end{eqnarray}
(Note that Eq.~\ref{eq:dsigma_db} corrects a misprint in Eq.~A8 of Ref.~\cite{Ford-14}.)
The derivatives of the response functions are given in terms of the Fermi phase by
\begin{equation}
\frac{\partial \hat{\chi}_{m}} {\partial a_{ }^{(\infty)}} 
 = \frac{\partial \hat{\chi}_{m}}{\partial \cos\left(\phi_{F}\right)} 
    \frac{\partial \cos\left(\phi_{F}\right)}{\partial a_{ }^{(\infty)}}
\end{equation}
\begin{equation}
\frac{\partial \hat{\chi}_{m}} {\partial b_{ }^{(\infty)}}
 = \frac{\partial \hat{\chi}_{m}}{\partial \cos\left(\phi_{F}\right)} 
    \frac{\partial \cos\left(\phi_{F}\right)}{\partial b_{ }^{(\infty)}}
\end{equation}
with partial derivatives of Eqs.~\ref{eq:chi} and~\ref{eq:Fermi-phase}
\begin{equation}
\frac{\partial \hat{\chi}_{m}} {\partial \cos\left(\phi_{F}\right)}  
 = -\frac{\cos(m+1)\phi_{F} - \cos(m-1)\phi_{F}}{\pi \sin\left(\phi_{F}\right)}
\end{equation}
\begin{equation}
\frac{\partial \cos\left(\phi_{F}\right)} {\partial a_{ }^{(\infty)}}
 = - \frac{1} {2 b_{ }^{(\infty)}}
\end{equation}
\begin{equation}
\frac{\partial \cos\left(\phi_{F}\right)} {\partial b_{ }^{(\infty)}} 
 = - \frac{\cos\left(\phi_{F}\right)} {2 b_{ }^{(\infty)}} \, .
\end{equation}
The derivatives of the recursion coefficients are given by~\cite{Horsfield-96-3}
\begin{equation}
  \frac{\partial a_n}{\partial \mu_m} = b_{n+1}  \sum_{j=0}^{n+1} \sum_{l=0}^n c_j^{n+1} c_l^n \delta_{l+j,m}
                                          -  b_n \sum_{j=0}^n \sum_{l=0}^{n-1} c_j^n c_l^{n-1} \delta_{l+j,m}
\end{equation}
\begin{equation}
  \frac{\partial b_n}{\partial \mu_m} = \frac{b_n}{2} 
       \left( \sum_{j=0}^n \sum_{l=0}^n c_j^n c_l^n \delta_{l+j,m} -
               \sum_{j=0}^{n-1} \sum_{l=0}^{n-1} c_j^{n-1} c_l^{n-1} \delta_{l+j,m} \right) \, .
\end{equation}
This set of partial derivatives is computed (i) in every self-consistency step to optimise the onsite-levels (Eq.~\ref{eq:BOP_SCF})
and (ii) in every force (Eq.~\ref{eq:Theta_tilde_BOP}) or torque calculation (Eq.~\ref{eq:torque}).

\section{Rotation matrices}
\label{sec:rot_mat}

The rotation matrices $R(\theta,\phi)$ in Eq.~\ref{eq:Hij_rot} are constructed from the polar and azimuthal angles $\theta$ and $\phi$ between the interatomic bond and the global coordinate system.
($\theta$ is the angle to the $xy$-plane and $\phi$ the angle to the $x$-axis in the $xy$-plane.)
For the orbital ordering in $H_{ij}$ of Eq.~\ref{eq:Hij}, the elements of the rotation matrix for $p$-orbitals are given by
\begin{eqnarray}
R(\theta,\phi)_{1,1} &=&  \cos(\theta)           \\ \nonumber
R(\theta,\phi)_{2,1} &=& -\sin(\theta)           \\ \nonumber
R(\theta,\phi)_{3,1} &=&  0.0                    \\ \nonumber
R(\theta,\phi)_{1,2} &=&  \cos(\phi)\sin(\theta) \\ \nonumber
R(\theta,\phi)_{2,2} &=&  \cos(\phi)\cos(\theta) \\ \nonumber
R(\theta,\phi)_{3,2} &=& -\sin(\phi)             \\ \nonumber
R(\theta,\phi)_{1,3} &=&  \sin(\phi)\sin(\theta) \\ \nonumber
R(\theta,\phi)_{2,3} &=&  \sin(\phi)\cos(\theta) \\ \nonumber
R(\theta,\phi)_{3,3} &=&  \cos(\phi)             \\ \nonumber
\end{eqnarray}
The matrix entries of the rotation matrix for $d$-orbitals are given by
\begin{eqnarray}
R(\theta,\phi)_{1,1} &=& \cos^2(\theta) - 1/2 \sin^2(\theta)                                   \\ \nonumber
R(\theta,\phi)_{2,1} &=& -\sqrt{3} \sin(\theta) \cos(\theta)                                   \\ \nonumber
R(\theta,\phi)_{3,1} &=& 0                                                                     \\ \nonumber
R(\theta,\phi)_{4,1} &=& \sqrt{3/4} \sin^2(\theta)                                             \\ \nonumber
R(\theta,\phi)_{5,1} &=& 0                                                                     \\ \nonumber
R(\theta,\phi)_{1,2} &=& \sqrt{3} \cos(\phi) \sin(\theta) \cos(\theta)                         \\ \nonumber
R(\theta,\phi)_{2,2} &=& \cos(\phi)(\cos^2(\theta) - \sin^2(\theta))                           \\ \nonumber
R(\theta,\phi)_{3,2} &=& -\sin(\phi) \cos(\theta)                                              \\ \nonumber
R(\theta,\phi)_{4,2} &=& -\cos(\phi) \sin(\theta) \cos(\theta)                                 \\ \nonumber
R(\theta,\phi)_{5,2} &=& \sin(\phi) \sin(\theta)                                               \\ \nonumber
R(\theta,\phi)_{1,3} &=& \sqrt{3} \sin(\phi) \sin(\theta) \cos(\theta)                         \\ \nonumber
R(\theta,\phi)_{2,3} &=& \sin(\phi) (\cos^2(\theta) - \sin^2(\theta))                          \\ \nonumber
R(\theta,\phi)_{3,3} &=& \cos(\phi) \cos(\theta)                                               \\ \nonumber
R(\theta,\phi)_{4,3} &=& -\sin(\phi) \sin(\theta) \cos(\theta)                                 \\ \nonumber
R(\theta,\phi)_{5,3} &=& -\cos(\phi) \sin(\theta)                                              \\ \nonumber
R(\theta,\phi)_{1,4} &=& (\cos^2(\phi) - \sin^2(\phi)) \sqrt{3/4} \sin^2(\theta)               \\ \nonumber
R(\theta,\phi)_{2,4} &=& (\cos^2(\phi) - \sin^2(\phi)) \sin(\theta) \cos(\theta)               \\ \nonumber
R(\theta,\phi)_{3,4} &=& -2 \sin(\phi)\cos(\phi) \sin(\theta)                                  \\ \nonumber
R(\theta,\phi)_{4,4} &=&  (\cos^2(\phi) - \sin^2(\phi)) (\cos^2(\theta) + 1/2 \sin^2(\theta))  \\ \nonumber
R(\theta,\phi)_{5,4} &=& -2 \sin(\phi)\cos(\phi) \cos(\theta)                                  \\ \nonumber
R(\theta,\phi)_{1,5} &=&  \sqrt{3} \sin(\phi)\cos(\phi) \sin^2(\theta)                         \\ \nonumber
R(\theta,\phi)_{2,5} &=&  2 \sin(\phi)\cos(\phi) \sin(\theta) \cos(\theta)                     \\ \nonumber
R(\theta,\phi)_{3,5} &=& (\cos^2(\phi) - \sin^2(\phi)) \sin(\theta)                            \\ \nonumber
R(\theta,\phi)_{4,5} &=& \sin(\phi)\cos(\phi) (\cos^2(\theta) + 1)                             \\ \nonumber
R(\theta,\phi)_{5,5} &=& (\cos^2(\phi) - \sin^2(\phi)) \cos(\theta)                            \\ \nonumber
\end{eqnarray}
Both rotation matrices become identity matrices for $\sin(\phi)=0$ and $\sin(\theta)=0$.
The rotation matrices for multiple orbital-types on one atom or for different orbitals on two interacting atoms are constructed by combinations of the above matrices.

\end{document}